\documentclass[twocolumn, trackchanges]{aastex61}
\usepackage{mathtools}
\usepackage{graphicx}
\usepackage{physics}
\usepackage{color}
\usepackage{xcolor}
\usepackage{hyperref}
\usepackage{amsmath}
\usepackage{afterpage}

\newcommand{\casa}{{\sc casa}}
\newcommand{\hi}{{\sc Hi}}

\shorttitle{HERA-19 Direction-dependent Effects}
\shortauthors{Kohn et al.}

\begin{document}

\title{\bf{The HERA-19 Commissioning Array: Direction Dependent Effects}}

\correspondingauthor{James E. Aguirre}
\email{jaguirre@sas.upenn.edu}

\author{Saul A. Kohn}
\affiliation{Department of Physics and Astronomy, University of Pennsylvania, Philadelphia, PA 19104 USA}

\author{James E. Aguirre}
\affiliation{Department of Physics and Astronomy, University of Pennsylvania, Philadelphia, PA 19104 USA}

\author{Paul La Plante}
\affiliation{Department of Physics and Astronomy, University of Pennsylvania, Philadelphia, PA 19104 USA}

\author{Tashalee S. Billings}
\affiliation{Department of Physics and Astronomy, University of Pennsylvania, Philadelphia, PA 19104 USA}

\author{Paul M. Chichura}
\affiliation{Department of Physics and Astronomy, University of Pennsylvania, Philadelphia, PA 19104 USA}

\author{Austin F. Fortino}
\affiliation{Department of Physics and Astronomy, University of Pennsylvania, Philadelphia, PA 19104 USA}

\author{Amy S. Igarashi}
\affiliation{Department of Physics and Astronomy, University of Pennsylvania, Philadelphia, PA 19104 USA}
\affiliation{Department of Astronomy, San Diego State University, San Diego, CA 92182 USA}

\author{Roshan K. Benefo}
\affiliation{Department of Physics and Astronomy, University of Pennsylvania, Philadelphia, PA 19104 USA}

\author{Samavarti Gallardo}
\affiliation{Department of Physics and Astronomy, University of Pennsylvania, Philadelphia, PA 19104 USA}
\affiliation{California State University of Los Angeles, 5151 State University Dr, Los Angeles, CA 90032 USA}

\author{Zachary E. Martinot}
\affiliation{Department of Physics and Astronomy, University of Pennsylvania, Philadelphia, PA 19104 USA}

\author{Chuneeta D. Nunhokee}
\affiliation{Department of Physics and Astronomy, University of Pennsylvania, Philadelphia, PA 19104 USA}
\affiliation{Department of Physics and Electronics, Rhodes University, PO Box 94, Grahamstown, 6140, South Africa}
\affiliation{Department of Astronomy, University of California, Berkeley, CA}

\author{Nicholas S. Kern} 
\affiliation{Department of Astronomy, University of California, Berkeley, CA}

\author{Philip Bull}
\affiliation{School of Physics \& Astronomy, Queen Mary University of London, Mile End Road, London E1 4NS, UK}

\author{Adrian Liu}
\affiliation{Department of Astronomy, University of California, Berkeley, CA}
\affiliation{Department of Physics and McGill Space Institute, McGill University, 3600 University Street, Montreal, QC H3A 2T8, Canada}


\author{Paul Alexander}
\affiliation{Cavendish Astrophysics, University of Cambridge, Cambridge, UK}

\author{Zaki S. Ali}
\affiliation{Department of Astronomy, University of California, Berkeley, CA}

\author{Adam P. Beardsley}
\affiliation{School of Earth and Space Exploration, Arizona State University, Tempe, AZ}
\affiliation{NSF Astronomy and Astrophysics Postdoctoral Fellow}

\author{Gianni Bernardi}
\affiliation{Department of Physics and Electronics, Rhodes University, PO Box 94, Grahamstown, 6140, South Africa}
\affiliation{INAF-Istituto di Radioastronomia, via Gobetti 101, 40129, Bologna, Italy}
\affiliation{SKA SA, 3rd Floor, The Park, Park Road, Pinelands, 7405, South Africa}

\author{Judd D. Bowman}
\affiliation{School of Earth and Space Exploration, Arizona State University, Tempe, AZ}

\author{Richard F. Bradley}
\affiliation{National Radio Astronomy Observatory, Charlottesville, VA}

\author{Chris L. Carilli}
\affiliation{Cavendish Astrophysics, University of Cambridge, Cambridge, UK}
\affiliation{National Radio Astronomy Observatory, Socorro, NM}

\author{Carina Cheng}
\affiliation{Department of Astronomy, University of California, Berkeley, CA}

\author{David R. DeBoer}
\affiliation{Department of Astronomy, University of California, Berkeley, CA} 

\author{Eloy de~Lera~Acedo}
\affiliation{Cavendish Astrophysics, University of Cambridge, Cambridge, UK} 

\author{Joshua S. Dillon}
\affiliation{Department of Astronomy, University of California, Berkeley, CA} 
\affiliation{NSF Astronomy and Astrophysics Postdoctoral Fellow}

\author{Aaron Ewall-Wice}
\affiliation{Jet Propulsion Laboratory, 4800 Oak Grove Dr, Pasadena, CA}
\affiliation{Dunlap Institute for Astronomy and Astrophysics, Toronto, Ontario, Canada}

\author{Gcobisa Fadana}
\affiliation{SKA-SA, Cape Town, South Africa}

\author{Nicolas Fagnoni}
\affiliation{Cavendish Astrophysics, University of Cambridge, Cambridge, UK} 

\author{Randall Fritz}
\affiliation{SKA-SA, Cape Town, South Africa}

\author{Steven R. Furlanetto}
\affiliation{Department of Physics and Astronomy, University of California, Los Angeles, CA}

\author{Brian Glendenning}
\affiliation{National Radio Astronomy Observatory, Socorro, NM}

\author{Bradley Greig}
\affiliation{ARC Centre of Excellence for All-Sky Astrophysics in 3 Dimensions
(ASTRO 3D), University of Melbourne, VIC 3010, Australia}
\affiliation{School of Physics, University of Melbourne, Parkville, VIC 3010, Australia}
 
\author{Jasper Grobbelaar}
\affiliation{SKA SA, 3rd Floor, The Park, Park Road, Pinelands, 7405, South Africa}

\author{Bryna J. Hazelton}
\affiliation{Department of Physics, University of Washington, Seattle, WA}
\affiliation{eScience Institute, University of Washington, Seattle, WA}

\author{Jacqueline N. Hewitt}
\affiliation{Department of Physics, Massachusetts Institute of Technology, Cambridge, MA}

\author{Jack Hickish}
\affiliation{Department of Astronomy, University of California, Berkeley, CA}

\author{Daniel C. Jacobs}
\affiliation{School of Earth and Space Exploration, Arizona State University, Tempe, AZ} 

\author{Austin Julius}
\affiliation{SKA SA, 3rd Floor, The Park, Park Road, Pinelands, 7405, South Africa}
 
\author{MacCalvin Kariseb}
\affiliation{SKA SA, 3rd Floor, The Park, Park Road, Pinelands, 7405, South Africa}
 
\author{Matthew Kolopanis} 
\affiliation{School of Earth and Space Exploration, Arizona State University, Tempe, AZ} 
 
\author{Telalo Lekalake}
\affiliation{SKA SA, 3rd Floor, The Park, Park Road, Pinelands, 7405, South Africa}

\author{Anita Loots}
\affiliation{SKA SA, 3rd Floor, The Park, Park Road, Pinelands, 7405, South Africa}

\author{David MacMahon}
\affiliation{Department of Astronomy, University of California, Berkeley, CA}
 
\author{Lourence Malan}
\affiliation{SKA SA, 3rd Floor, The Park, Park Road, Pinelands, 7405, South Africa}
 
\author{Cresshim Malgas}
\affiliation{SKA SA, 3rd Floor, The Park, Park Road, Pinelands, 7405, South Africa}

\author{Matthys Maree}
\affiliation{SKA SA, 3rd Floor, The Park, Park Road, Pinelands, 7405, South Africa}

\author{Nathan Mathison}
\affiliation{SKA SA, 3rd Floor, The Park, Park Road, Pinelands, 7405, South Africa}

\author{Eunice Matsetela}
\affiliation{SKA SA, 3rd Floor, The Park, Park Road, Pinelands, 7405, South Africa} 

\author{Andrei Mesinger}
\affiliation{Scuola Normale Superiore, Pisa, Italy}

\author{Miguel F. Morales}
\affiliation{Department of Physics, University of Washington, Seattle, WA} 

\author{Abraham R. Neben}
\affiliation{Department of Physics, Massachusetts Institute of Technology, Cambridge, MA}

\author{Bojan Nikolic}
\affiliation{Cavendish Astrophysics, University of Cambridge, Cambridge, UK} 

\author{Aaron R. Parsons}
\affiliation{Department of Astronomy, University of California, Berkeley, CA}

\author{Nipanjana Patra}
\affiliation{Department of Astronomy, University of California, Berkeley, CA}

\author{Samantha Pieterse}
\affiliation{SKA SA, 3rd Floor, The Park, Park Road, Pinelands, 7405, South Africa}
 
\author{Jonathan C. Pober}
\affiliation{Department of Physics, Brown University, Providence, RI} 

\author{Nima Razavi-Ghods}
\affiliation{Cavendish Astrophysics, University of Cambridge, Cambridge, UK}

\author{Jon Ringuette}
\affiliation{Department of Physics, University of Washington, Seattle, WA}

\author{James Robnett}
\affiliation{National Radio Astronomy Observatory, Socorro, NM}

\author{Kathryn Rosie}
\affiliation{SKA SA, 3rd Floor, The Park, Park Road, Pinelands, 7405, South Africa}

\author{Raddwine Sell}
\affiliation{SKA SA, 3rd Floor, The Park, Park Road, Pinelands, 7405, South Africa}

\author{Craig Smith}
\affiliation{SKA SA, 3rd Floor, The Park, Park Road, Pinelands, 7405, South Africa}

\author{Angelo Syce}
\affiliation{SKA SA, 3rd Floor, The Park, Park Road, Pinelands, 7405, South Africa}

\author{Max Tegmark}
\affiliation{Department of Physics, Massachusetts Institute of Technology, Cambridge, MA} 

\author{Nithyanandan Thyagarajan}
\affiliation{School of Earth and Space Exploration, Arizona State University, Tempe, AZ}
\affiliation{National Radio Astronomy Observatory, Socorro, NM}
\affiliation{Jansky Fellow}

\author{Peter K.~G. Williams}
\affiliation{American Astronomical Society, Washington, DC 20006 USA}
\affiliation{Center for Astrophysics | Harvard \& Smithsonian, Cambridge, MA 02138 USA}


\author{Haoxuan Zheng}
\affiliation{Department of Physics, Massachusetts Institute of Technology, Cambridge, MA}

%
%
%

\begin{abstract}
  Foreground power dominates the measurements of interferometers that seek a
  statistical detection of highly-redshifted \hi\ emission from the Epoch of
  Reionization (EoR).  The chromaticity of the instrument creates a boundary in
  the Fourier transform of frequency (proportional to $k_\parallel$) between
  spectrally smooth emission, characteristic of the strong synchrotron
  foreground (the ``wedge''), and the spectrally structured emission from \hi\
  in the EoR (the ``EoR window'').  Faraday rotation can inject spectral
  structure into otherwise smooth polarized foreground emission, which through
  instrument effects or miscalibration could possibly pollute the EoR window.
  For instruments pursuing a ``foreground avoidance'' strategy of simply
  measuring in the EoR window, and not attempting to model and remove
  foregrounds, as is the plan for the first stage of the Hydrogen Epoch of
  Reionization Array (HERA), characterizing the intrinsic instrument
  polarization response is particularly important.  Using data from the HERA
  19-element commissioning array, we investigate the polarization response of
  this new instrument in the power spectrum domain.  We perform a simple
  image-based calibration based on the unpolarized diffuse emission of the
  Global Sky Model, and show that it achieves qualitative redundancy between the
  nominally-redundant baselines of the array and reasonable amplitude accuracy.
  We construct power spectra of all fully polarized coherencies in all
  pseudo-Stokes parameters, and discuss the achieved isolation of foreground
  power due to the intrinsic spectral smoothness of the foregrounds, the
  instrument chromaticity, and the calibration.  We compare to simulations based
  on an unpolarized diffuse sky model and detailed electromagnetic simulations
  of the dish and feed, confirming that in Stokes I, the calibration does not
  add significant spectral structure beyond that expected from the
  interferometer array configuration and the modeled primary beam response.
  Further, this calibration is stable over the 8 days of observations
  considered.  Excess power is seen in the power spectra of the linear
  polarization Stokes parameters which is not easily attributable to leakage via
  the primary beam, and results from some combination of residual calibration
  errors and actual polarized emission.  Stokes V is found to be highly
  discrepant from the expectation of zero power, strongly pointing to the need
  for more accurate polarized calibration.
\end{abstract}


\section{Introduction}
\label{sec:intro}

Many low-frequency (50 -- 200\,MHz) radio interferometers
(e.g. LOFAR\footnote{\url{www.lofar.org}},
MWA\footnote{\url{www.mwatelescope.org}},
PAPER\footnote{\url{eor.berkeley.edu}},
HERA\footnote{\url{www.reionization.org}}) around the world are seeking to
detect brightness-temperature fluctuations of neutral hydrogen during the Epoch
of Reionization (EoR; for an overview see \citet{Furlanetto06}).  Such a
detection is predicted to be rich in information about the astrophysics and
cosmology of the high-redshift ($\sim 7 < z < 14$) Universe.  The {\sc Hi}
brightness-temperature fluctuations are not only intrinsically faint but also
hidden by foreground emission. Foreground emission, predominantly in the form of
galactic and extragalactic synchrotron emission, is many orders of magnitude
more powerful than the cosmological signal \citep[e.g.][]{Bernardi09, Pober13,
  Dillon14}.

Most foreground emission is due to synchrotron emission, which is spectrally
smooth. The instrumental response of an interferometer is inherently chromatic,
and the cosmological signal is spectrally structured. In sum, this leads to the
property that Fourier transforming the interferometric measurement along the
frequency axis delineates a boundary in the $\mathbf{k}$-space between the
spectrally smooth foregrounds (in the ``wedge'') and the cosmological {\sc Hi}
signal (in the ``EoR window'') \citep{Datta.10, Morales.12, Parsons.12a,
  Parsons.12b, Trott.12, Vedantham.12, Pober13, Thyagarajan.13, Pober.14,
  Liu.14a, Liu.14b, Dillon.15a, Dillon.15b, Nithya.15b, Nithya.15a}.  Thermal
noise is present throughout this space, and dominates the EoR window in any
single observation.  Detection of the EoR thus requires long observing seasons,
precision calibration, and suppression of instrument systematics.

The cosmological {\sc Hi} signal is strongly unpolarized
\citep{Mishra17}. However, polarized synchrotron radiation represents a
potential foreground contaminant capable of leaking into the EoR window. At low
frequencies, Faraday rotation can impart significant spectral structure to
polarized emission \citep[e.g.][]{Moore13}. This polarized signal is able to
`leak' into unpolarized measurements due to miscalibration and instrumental
effects \citep{Carozzi.09, Geil.11, Moore13, Asad15, Asad.16, Kohn16,
  Nunhokee.17}, contaminating the EoR window.

It is important to constrain intrinsic and leaked polarized signal for any {\sc
  Hi} intensity mapping experiment. The objective of this paper is an
exploration of eight nights of data from the Hydrogen Epoch of Reionization
Array (HERA) 19-element commissioning array, coupled with simulations of the
instrument, in order to characterize the polarized response of this
interferometer.  One of the more difficult features of an interferometer to
characterize is the frequency- and direction-dependent polarized antenna
response, which is important for characterizing polarized-to-unpolarized leakage
in the wedge/window paradigm \citep{Moore17,Nunhokee.17,Martinot18}.  In this
work, we were primarily sensitive to leakage in the unpolarized-to-polarized
direction. Due to the symmetry of leakage modes (elaborated upon in
Section~\ref{sec:leak}), this still represents a useful constraint on the future
problem of polarized-to-unpolarized leakage contaminating the EoR signal.  While
we use cosmological power spectra as a diagnostic of the data, the goal of this
paper is not to obtain new upper limits on the EoR power spectrum, but simply to
integrate deep enough to test models of the instrument's spectral response
against simulations.

This work is organized as follows: in Section~\ref{sec:leak} we review the
theory behind polarization leakage into unpolarized signal and present the
polarized primary beam model for the HERA commissioning array. In
Section~\ref{sec:obs} we describe the HERA data that we used, its calibration,
and reduction to power spectra. We present our results, and discuss the
implications in Section~\ref{sec:results}, and conclude in Section~\ref{conc}.

We assume the cosmological parameters reported by \cite{Planck.16} throughout.

\section{Leakage Modes}
\label{sec:leak}

A radio interferometer measures correlations of voltages. Viewed in
transmission, a dipole arm of antenna $i$ radiates a far-field electric field
pattern
\begin{equation}
\vec{E}_{i}(\hat{s}, \nu) = E_{i,\theta}(\nu)\hat{\theta} + E_{i,\phi}(\nu)\hat{\phi}
\end{equation}
where $(\hat{\theta},\hat{\phi})$ define an orthogonal coordinate system on the
sphere. These far-field beam patterns, by the reciprocity theorem, define the
response of the feed to an electric field from infinity in the direction
$(\theta,\phi)$.

We may choose to express the electric field response in Right Ascension and
Declination basis (unit vectors $\hat{e}_{\alpha}$, $\hat{e}_{\delta}$),
allowing us to express the coherency tensor field
\begin{multline}
\mathcal{C} =  \langle E_{\delta}^* E_{\delta} \rangle \hat{e}_{\delta} \otimes \hat{e}_{\delta} 
					+  \langle E_{\alpha}^* E_{\delta} \rangle \hat{e}_{\alpha} \otimes \hat{e}_{\delta} \\
					+  \langle E_{\delta}^* E_{\alpha} \rangle \hat{e}_{\delta} \otimes \hat{e}_{\alpha}
					+  \langle E_{\alpha}^* E_{\alpha} \rangle \hat{e}_{\alpha} \otimes \hat{e}_{\alpha} 
\end{multline}
where we have dropped the explicit $(\hat{s}, \nu)$ dependence of the fields.
By definition, the coherency field is specified by the Stokes parameters
\begin{equation}
\mathcal{C} = \begin{pmatrix}
I(\hat{s}, \nu) + Q(\hat{s}, \nu) & U(\hat{s}, \nu) - iV(\hat{s}, \nu) \\
U(\hat{s}, \nu) + iV(\hat{s}, \nu) & I(\hat{s}, \nu) - Q(\hat{s}, \nu) \\
\end{pmatrix}.
\end{equation}

Each polarized feed $p$ of antenna $i$ responds to incident radiation from
direction $(\hat{\theta},\hat{\phi})$ with a complex vector antenna pattern
\begin{equation}
\vec{A}^p_i(\hat{s},\nu) = A_{i,\theta}^p(\hat{s},\nu)\hat{\theta} + A_{i,\phi}^p(\hat{s},\nu)\hat{\phi}.
\end{equation}
The antenna patterns can be written as components of a direction-dependent Jones
matrix for a dipole feed $i$ with arms $p$ and $q$:
\begin{equation}
\mathcal{J}_i = 
\begin{pmatrix}
A_{i,\theta}^p(\hat{s},\nu) & A_{i,\phi}^p(\hat{s},\nu) \\
A_{i,\theta}^q(\hat{s},\nu) & A_{i,\phi}^q(\hat{s},\nu)
\end{pmatrix}.
\end{equation}
We can then express the fully-polarized visibility equation for the correlation of feeds $i$ and $j$ as 
\begin{equation}
\mathcal{V}_{ij} = \int \mathcal{J}_i\mathcal{C}\mathcal{J}_j^{\dagger} \exp(-2\pi i \nu \vec{b}\cdot\hat{s} / c){\rm d}\Omega = \begin{pmatrix}
V^{nn}_{ij} & V^{ne}_{ij}\\
V^{en}_{ij} & V^{ee}_{ij}\\
\end{pmatrix}
\end{equation}
where we have denoted dipole arms $p$ and $q$ as $n$ and $e$, representing a
configuration where the arms are oriented along the North-South and East-West
directions, respectively.

Unless $\mathcal{J}$ is both diagonal and has, at any given point on the sphere,
equal diagonal elements, there will be mixing or ``leaking'' of different Stokes
parameters together into each element of $\mathcal{V}$ in a direction dependent
way \citep{Geil.11,Smirnov.11.1,Smirnov.11.2,Nunhokee.17}.

\subsection{Direction-Dependent Leakage}
\label{subsec:DD-Leak}

The cosmological signal of interest for 21cm cosmology studies is effectively
unpolarized, and we therefore use the pseudo-Stokes\footnote{We use
  ``pseudo-Stokes'' to refer to Stokes parameters formed from visibilities
  throughout this work, to distinguish from true ``Stokes parameters'' defined
  in the image domain by the IEEE \citep{Ludwig.73, vanStraten.10}.} I
visibility to measure it \citep[e.g.][]{Moore13}; this is defined
  $V^{I} = V^{nn} + V^{ee}$, which is the trace of $\mathcal{V}$:
\begin{multline}
V^I_{ij}(\nu) = \rm{Tr}(\mathcal{V}_{ij}) = \int \rm{Tr}(\mathcal{J}_i\mathcal{C}\mathcal{J}_j^{\dagger}) \exp(-2\pi i \nu \vec{b} \cdot \hat{s} / c)  {\rm d}\Omega \\
= \int (\mathcal{M}_{00}I + \mathcal{M}_{01}Q + \mathcal{M}_{02}U + \mathcal{M}_{03}V)\exp(-2\pi i \nu \vec{b}\cdot\hat{s} / c){\rm d}\Omega 
\end{multline}
where $I$, $Q$, $U$ and $V$ are the true Stokes sky and are functions of
direction and frequency, and $\mathcal{M}_{ab}(\hat{s},\nu)$ are the
instrumental Mueller matrix elements:
\begin{equation}
\mathcal{M}_{ab}(\hat{s},\nu) = {\rm Tr}(\sigma_a \mathcal{J} \sigma_b \mathcal{J}^{\dagger})
\end{equation}
and $\sigma_i$ are the Pauli matrices (where the indices are reordered from the
quantum mechanical convention to an order which gives the ordering of the Stokes
vector as ($I$, $Q$, $U$, $V$); see, e.g., \citealt{Shaw.15.1}).

\begin{figure*}
\centering
\includegraphics[scale=0.4]{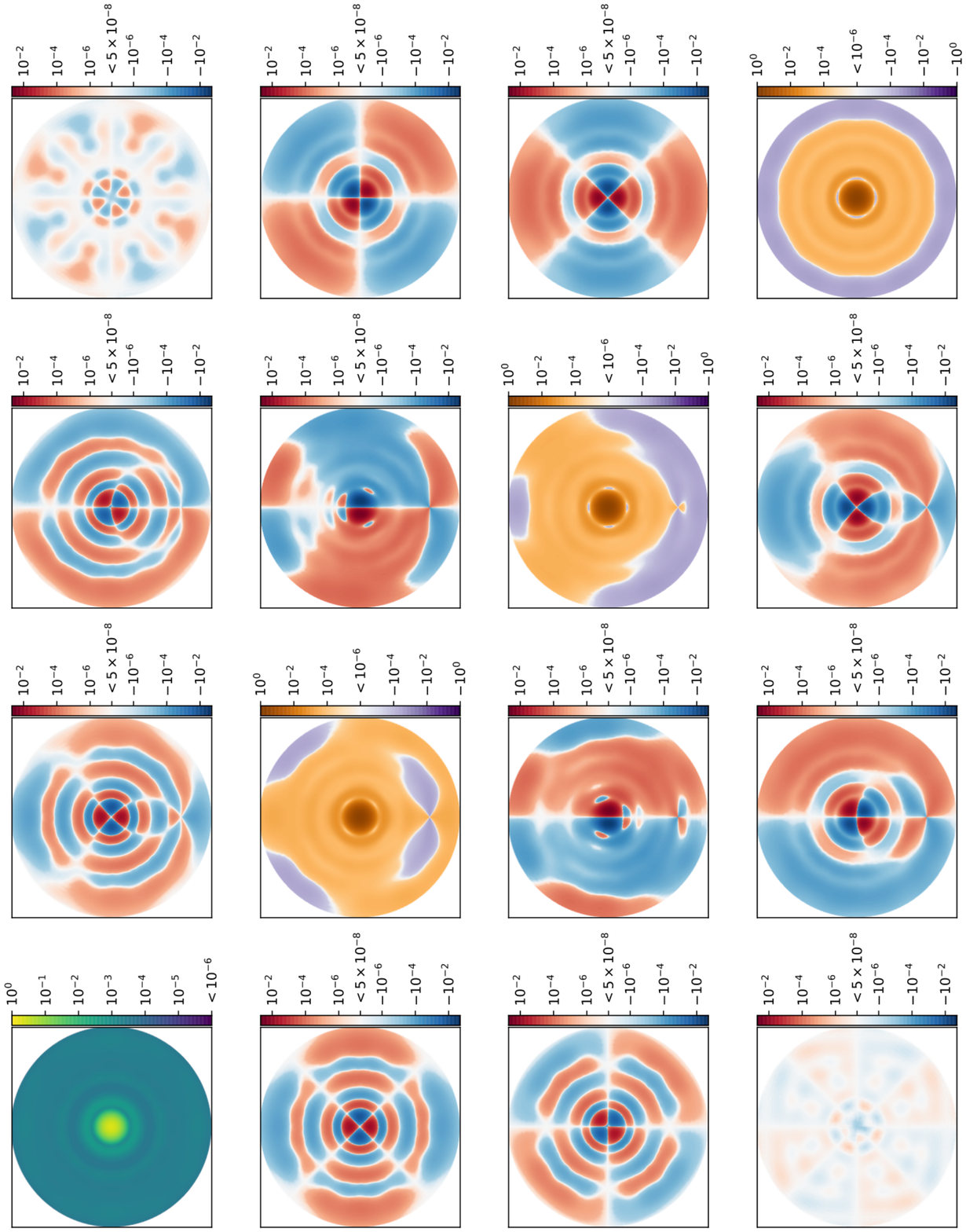}
\par\noindent\rule{0.8\textwidth}{0.4pt}
\includegraphics[scale=0.4]{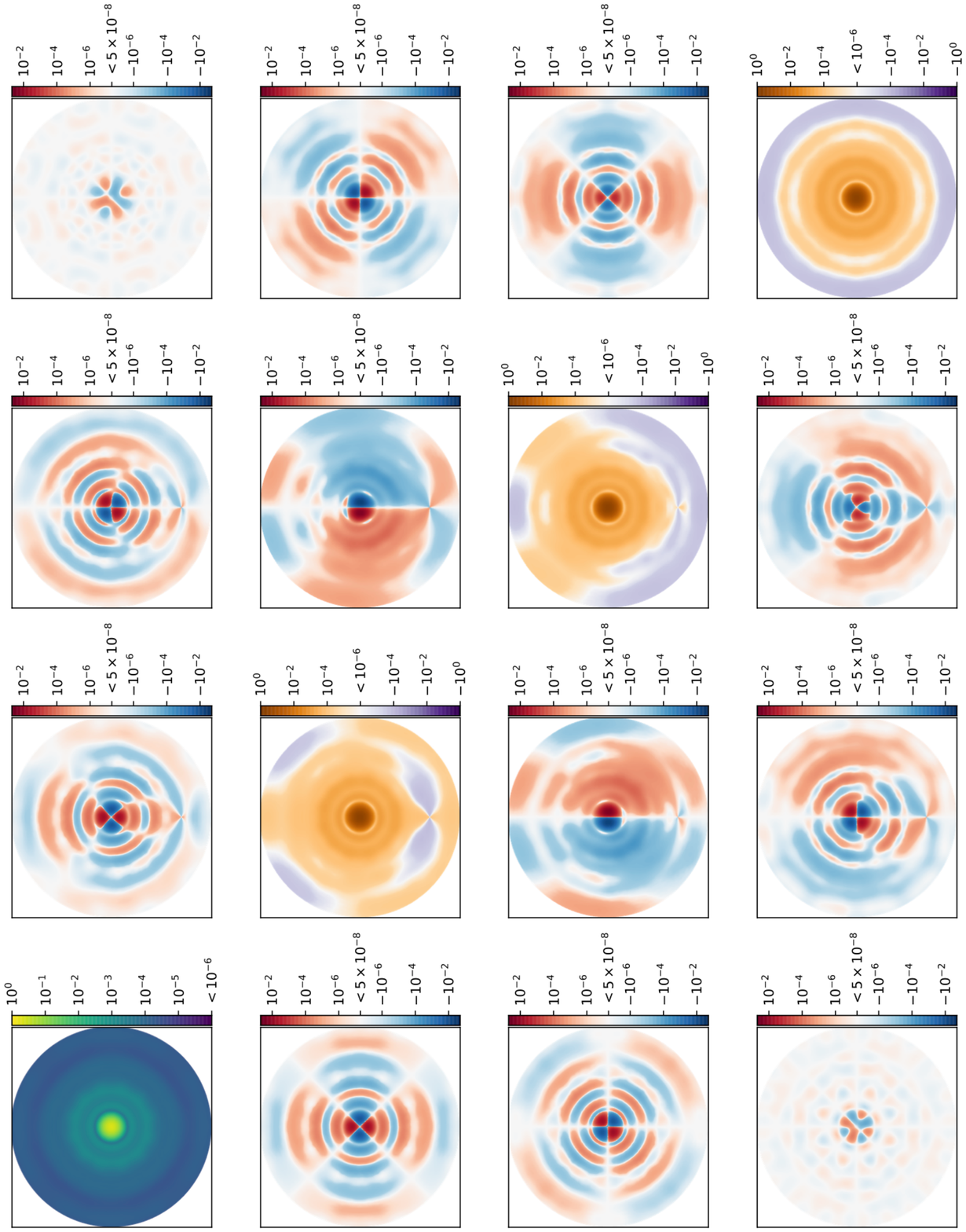}
\caption{Simulations of the instrumental direction dependent Mueller matrix at
  120 MHz and 160 MHz (\textit{above} and \textit{below}, respectively)
  projected into the RA, Dec basis. Color scales for frequencies are relative to
  the peak of $\mathcal{M}_{00}$ (which itself is normalized to 1 at zenith). To
  account for the wide variety of dynamic ranges required to show detail, we use
  separate color maps for $\mathcal{M}_{00}$, diagonal, and off-diagonal
  terms. The off-diagonal terms are 2- to 8-orders of magnitude less than the
  diagonal terms. For a key to these matrices, see Equation~\ref{eq:Mab}.  }
\label{fig:mueller}
\end{figure*}

We simulated the HERA feed, faceted parabolic dish and analog signal chain using
CST\footnote{\url{www.cst.com}} to generate the complex $\vec{E}$-field
receptivity patterns, as described in \cite{Fagnoni.16} (also see public
\href{http://reionization.org/wp-content/uploads/2013/03/HERA_memo_21_CST_simulation_of_HERA_and_comparison_with_measurements.pdf}{\underline{HERA
    Memo \#21}}), and then formed $\mathcal{J}$ and $\mathcal{M}$ as described
above. Examples of $\mathcal{M}_{ij}$ at 120 MHz and 160 MHz (our low and high
bands of interest; see Section~\ref{subsec:cal}) are shown in
Figure~\ref{fig:mueller}, projected in the RA/Dec basis. Note that this basis
has a singularity at the South Pole, leading to wide-field asymmetries in Q and
U. Due to the large spread in dynamic ranges between $\mathcal{M}_{00}$, other
diagonal terms, and off-diagonal terms, we use separate color maps for each. All
of the dynamic ranges are normalized to the peak of $\mathcal{M}_{00}$, which is
1 at zenith. The off-diagonal terms are 2- to 8-orders of magnitude less than
the diagonal terms.

The key for these matrices are the mappings of Stokes parameters into
pseudo-Stokes visibilities, following
\begin{equation}
\mathcal{M}_{ab}(\hat{s},\nu) =
\begin{pmatrix}
I \rightarrow V^I & I \rightarrow V^Q & I \rightarrow V^U & I \rightarrow V^V\\
Q \rightarrow V^I  & Q \rightarrow V^Q & Q \rightarrow V^U & Q \rightarrow V^V\\
U \rightarrow V^I  & U \rightarrow V^Q & U \rightarrow V^U & U \rightarrow V^V\\
V \rightarrow V^I  & V \rightarrow V^Q & V \rightarrow V^U & V \rightarrow V^V\\
\end{pmatrix}
\label{eq:Mab}
\end{equation}
where pseudo-Stokes visibilities are formed as:
\begin{equation}
\left(\begin{array}{c}
V^{I}\\
V^{Q}\\
V^{U}\\
V^{V}\end{array} \right)
= \frac{1}{2}
\left( \begin{array}{cccc}
1 & 0 & 0 & 1 \\
1 & 0 & 0 & -1 \\
0 & 1 & 1 & 0 \\
0 & -i & i & 0 \end{array} \right) 
\left(\begin{array}{c}
V^{nn}\\
V^{ne}\\
V^{en}\\
V^{ee}\end{array} \right) .
\label{eq:pseudo-stokes}
\end{equation}

\begin{figure}
\centering
\hspace{-0.5cm}
\includegraphics[scale=0.6]{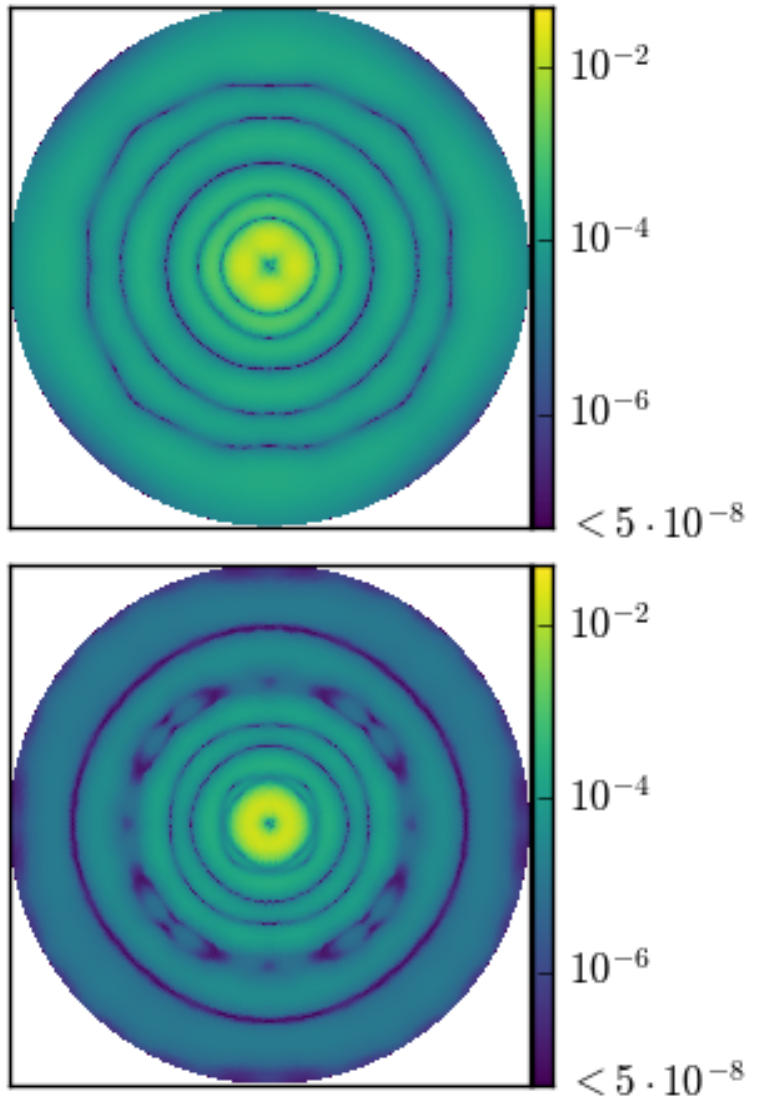}
\caption{The magnitude of the linear polarization leakage beam given by
    $\mathcal{M}_p = \sqrt{\mathcal{M}_{IQ}^2 + \mathcal{M}_{IU}^2}$, or the
    middle two entries of the top row of Figure \ref{fig:mueller}, at 120 MHz
    and 160 MHz (\textit{above} and \textit{below}, respectively).
}
\label{fig:lin_pol_beam}
\end{figure}

At low frequencies and the large scales probed by many low frequency
interferometers, Stokes I is extremely bright compared to the other Stokes
parameters \citep{Bernardi.09, Bernardi.10, Jelic.14, Jelic.15, Asad15, Kohn16,
  Lenc.17, Moore17}. Moreover, only a few polarized point sources have been
observed at frequencies below 300 MHz \citep{Bernardi.13, Asad.16,
  Lenc.17}. \cite{Farnes.14} showed evidence for systematic depolarization of
steep-spectrum point sources towards low frequencies, causing low polarization
fractions ($\ll 1\%$) below 300 MHz.

The ``linear polarization leakage beam'' is shown for the two central
frequencies of this analysis in Figure \ref{fig:lin_pol_beam}.  This quantity is
the magnitude of the spin-2 function $\mathcal{M}_{IQ} + i \mathcal{M}_{IU}$ and
represents the amplitude of the direction-dependent leakage of Stokes Q and U
into I.

These factors make the first row of $\mathcal{M}$, which represents
$I\rightarrow V^I,\,V^Q,\,V^U,\,V^V$, the most interesting for low-frequency
polarized power spectra, since with limited calibration we can expect leakage
from Stokes I into the other Stokes parameters to dominate over Stokes Q, U, and
V emission alone.

We produced simulations $\mathcal{V}$ using our fully-polarized formalism for
the HERA-19 commissioning array, described below, using an unpolarized model of
the low frequency sky from the Global Sky Model \citep[GSM;][]{GSM.08, pygsm,
  GSM.17} at the appropriate R.A. range to match our observations. These
simulations are based on the same source code as \citet{Martinot18}.

Forming power spectra from these visibilities allowed for a comparison of our
data to a ``leakage only'' regime. We discuss the process for forming power
spectra in Section~\ref{subsec:pspec}, and the simulated power spectra are shown
in comparison to those from data in Section~\ref{sec:results}.

\subsection{Direction-Independent Leakage}
\label{subsec:DI-Leak}

In addition to the mixing of Stokes parameters due to the primary beam, it is
possible to mix them in a direction independent way. Calibration errors are
capable of leaking signal between pseudo-Stokes visibilities independent of the
sky \citep{TMS}. Again focusing on the $\{V^I,\,V^Q,\,V^U,\,V^V\}$ components of
this leakage, we have:
\begin{itemize}
\item $V^I \rightarrow V^Q$ occurs through errors in calibrating the complex
  voltage gain factors for each dipole arm.
\item $V^I \rightarrow V^U$ occurs through the sum of off-diagonal gain terms
  (\textit{D}-terms; the receptivity of dipole arm ``n'' to an electric field
  vector aligned with arm ``e'' and vice versa).
\item $V^I \rightarrow V^V$ occurs through the difference in \textit{D}-terms
  between two feeds.
\end{itemize}

We detail how we obtain the direction-independent Jones terms in the next
section.

\section{Observations and Reduction}
\label{sec:obs}

\begin{figure}
\centering
\hspace{-0.5cm}
\includegraphics[scale=0.6]{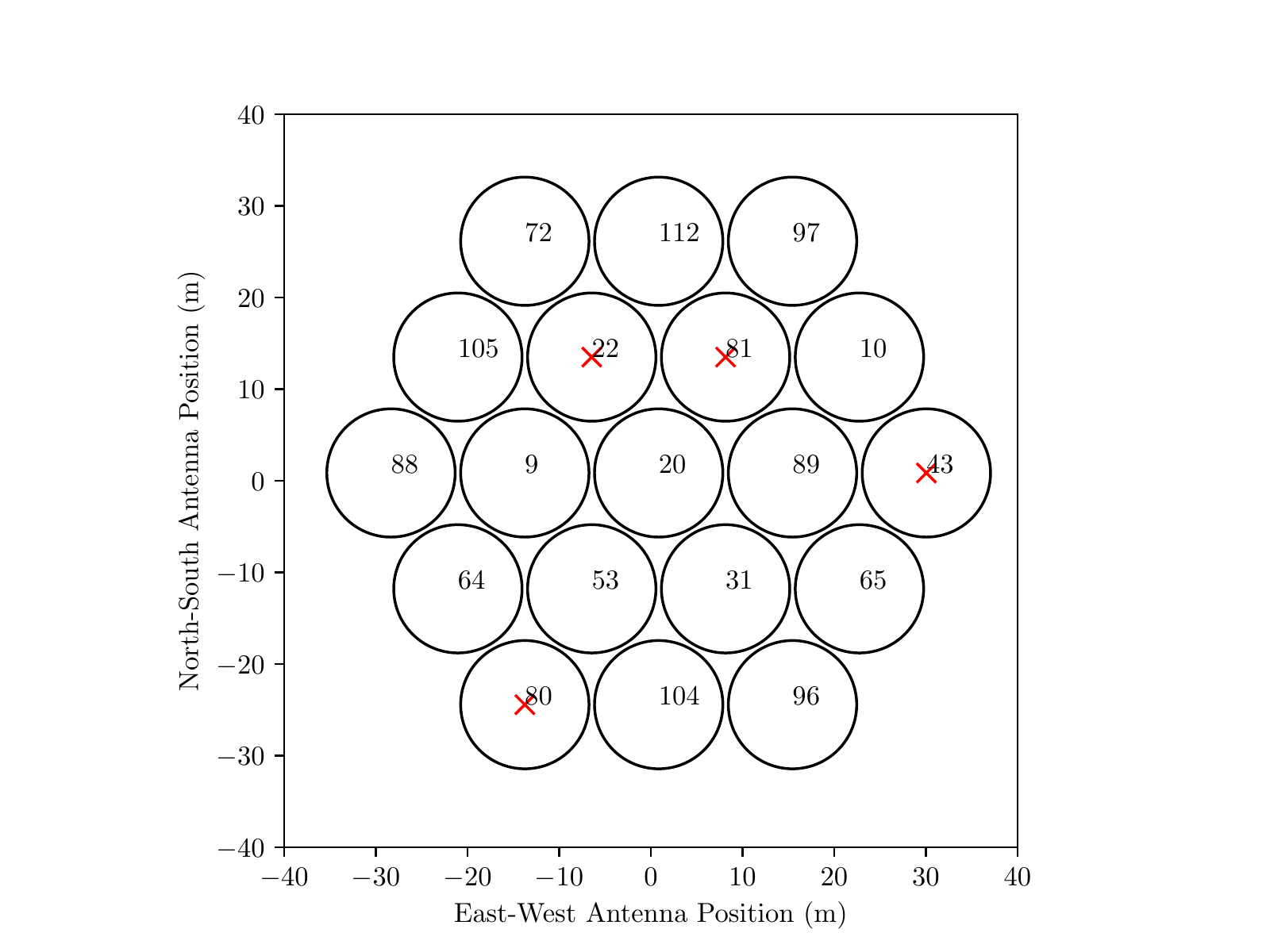}
\caption{The configuration of the HERA-19 array. The perimeter of each dish is
  shown as a circle. A red ``X'' marks antennas that were identified during
  preprocessing and calibration as malfunctioning and were excluded from further
  analysis.}
\label{fig:antpos}
\end{figure}

In this work we used eight nights of observations from the HERA-19 commissioning
array. HERA is a low-frequency interferometer composed of 14\,m-diameter dishes
arranged in a close-packed hexagonal array of 14.7\,m spacing. The commissioning
array consists of nineteen dishes (see Figure~\ref{fig:antpos}); HERA is being
constructed in staged build-outs, and upon completion will consist of 350 dishes
in a fractured hexagon configuration \citep[see][]{DillonParsons16, deBoer17}. A
feed cage containing two dipole feeds (recycled from the PAPER array, see
\citealt{Parsons.10}), oriented in North-South and East-West directions, was
suspended above each dish \citep{Neben.16,Ewall-Wice.16,Thyagarajan.16}.

HERA only observes in drift-scan mode. The observations we used were eight
nights, from Julian Date (JD) 2457548 to 2457555; LSTs 10.5 -- 23 hr. Drift-scan
visibilities were recorded every 10.7 seconds for 1024 evenly-spaced channels
across the 100-200\,MHz bandwidth. These data were divided into {\sc miriad}
data sets roughly 10 minutes long. A night's observation lasted 12 hours in
total (6pm to 6am South African Standard Time; SAST); of these we used the
central 10 hours, to avoid the sun. A summary of the instrument and observation
parameters is given in Table~\ref{tab:params}.

\begin{table}
\centering
\caption{Observational parameters used for this study.}
\begin{tabular}{lc}
\hline
Parameter & Value \\
\hline
Array location & 30:43:17.5$^\circ$ S, 21:25:41.9$^\circ$ E \\
JD range & 2457548 -- 2457555 \\
LST Range & 10.6 -- 22.6 hrs \\
Frequency range & 115 -- 185 MHz \\
Frequency resolution & 97.6 kHz \\
Integration time & 10.7 s\\
Element diameter & 14.0 m\\
Number of elements & 15 \\
Shortest baseline & 14.6 m \\
Longest baseline & 58.4 m \\
At 150 MHz: & \\
Primary beam FWHM & $9^{\circ}$ \\
Synthesized beam FWHM & 2\arcdeg \\
SEFD per element  & $\sim 5800$ Jy \\

\hline
\end{tabular}
\label{tab:params}
\end{table}

\subsection{RFI Excision and Flagging}

\begin{figure}
\centering
\includegraphics[scale=0.6]{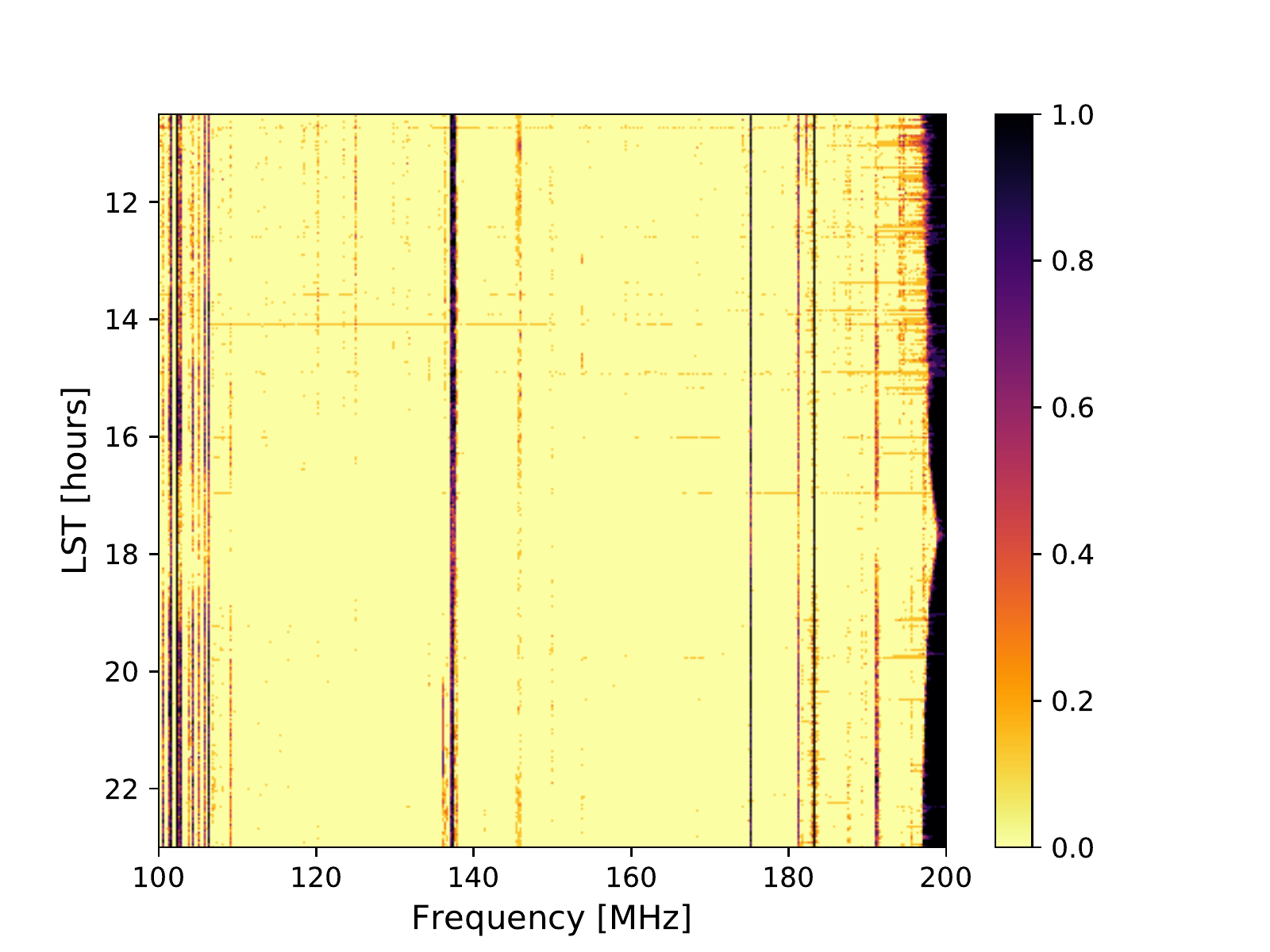}
\caption{Fractional RFI flag occupancy per time and frequency over the eight
  days of observations. RFI was flagged on a per-(time,frequency) sample basis.}
\label{fig:rfi}
\end{figure}

To identify samples contaminated by radio frequency interference (RFI), a
two-dimensional median filter in time and frequency was applied to the
visibility data to smooth out high pixel-to-pixel variations, and remove
significant outliers that were likely unphysical. The variance of the resulting
data was computed, and points with a $z$-score greater than 6 (i.e., points
where the value is more than 6$\sigma$ away from the mean) were flagged as
initial seeds for RFI extraction. A two-dimensional watershed algorithm was
applied using these seeds as starting points, enlarging the regions of
RFI-contamination to neighboring pixels with $z$-scores greater than 2, until
all such pixels were flagged. Figure~\ref{fig:rfi} shows the fractional RFI flag
occupancy per time (displayed in LST) and frequency across the 8 days of
observations. The majority of the band is relatively clear of RFI. Some clear
features are: the FM radio band (below 110 MHz), ORBCOMM satellite
communications (137 MHz), an ISS downlink (150 MHz) and VHF TV channels (above
170 MHz)\footnote{For an extended discussion of RFI as seen by HERA, see the
  public
  \href{http://reionization.org/wp-content/uploads/2013/03/HERAMemo19_HERA_dish_RFI.pdf}{\underline{HERA
      Memo \#19}}}.  The galaxy, when transiting zenith at LST$\approx$17.75
hours, is so bright that it appears to degrade our ability to flag RFI.

Four antennas were identified during the commissioning as having anomalous
behavior. These are marked with red ``X''s in Figure~\ref{fig:antpos} and were
omitted from further analysis.  Before calibration, we manually flagged the
edges of the band (below 110 MHz and above 190 MHz), where spectral behavior is
dominated by the high and low pass filtering in the HERA signal chain
\citep{deBoer17}.

\subsection{Calibration}
\label{subsec:cal}

HERA is designed to be calibrated using redundant calibration techniques
\citep{DillonParsons16}, but for this preliminary view of HERA commissioning
data, we used image-based calibration. Future studies with deeper integrations
targeting EoR detections will take advantage of redundancy to obtain more
precise calibration solutions \citep{deBoer17}. We used the \casa\ \citep{casa}
package for calibration, taking advantage of its CLEAN, {\tt gaincal} and {\tt
  bandpass} functions.  We first converted from HERA's native {\sc miriad} to a
{\sc uvfits} file format using {\sc pyuvdata} \citep{pyuvdata}; this could then
be ingested by \casa.

The brightest calibration sources near the Dec -30\arcdeg\ stripe -- for
example, those used in previous PAPER analyses like Pictor A \citep{jacobs13b}
-- were not available for this observing window (10.5 - 23 h RA), and the few
long baselines in the array available made calibration using individual fainter
point sources difficult.  We therefore developed a calibration method using the
Galactic Center (GC; taken to be at $\alpha, \delta$ = 17h 45m 40s, -29d 0m 28s)
as modeled by the GSM. Specifically, we selected four minutes of data centered
at the transit of the GC to use for calibration.  The visibilities were phased
from drift-scan mode to a single phase center chosen to be the LST at the start
of the observations.  Because this phasing imperfectly approximates the tracking
telescope assumed by \casa, the length of the observation was chosen to minimize
the effects of beam-dependent gain variations as the GC transited.  The
calibration was done as a two-step process.  First, we built an initial model
with the GC as an unpolarized, flat-spectrum source with flux density scaled to
a reference point of 1 Jy.  This allowed us to solve for the large antenna based
delay terms using {\tt gaincal} ({\tt gaintype='K'}; typically 10's of ns) and a
complex bandpass using {\tt bandpass}.  A single solution was obtained for the 4
minute observation for both calibration types, and for the bandpass a solution
was obtained for each unflagged $\sim$100 kHz channel, resulting in a complex,
frequency-dependent gain for each feed. With this first calibration in place,
the second step was to interactively CLEAN the image to obtain a more accurate
model of the GC extended structure.  We still assumed an unpolarized source, but
allowed multiple components within a two degree radius centered on the GC.  This
was followed by a second round of delay and bandpass calibration to the
multi-component extended model, completely analogous to first round of
calibration.  At this point, an overall frequency-dependent amplitude was
required to scale the gains from the arbitrary 1 Jy normalization.  For this we
used our simulation of the GC from the GSM (converted to units of Jy) to
determine a single, spectrally-smooth function for all antennas to make the
spectrum of the observations match the simulations.

Clearly, this is an incomplete calibration model.  The assumption that the GC is
unpolarized is probably adequate, due to the large optical depth towards the GC
\citep{Oppermann.12} resulting in near-complete depolarization in the plane of
the Galaxy \citep{Wolleben.06}.  Moreover, we expect significant beam
depolarization due to the large solid angle of the synthesized beam (see Table
\ref{tab:params}).  Other assumptions are less obviously correct.  The GC
structure is only partially modeled, and we have assumed the GSM provides an
accurate calibration.  We have also assumed that the direction-independent Jones
matrix is diagonal.  Although in principal \casa\ is capable of solving for
\textit{D}-terms using the {\tt polcal} task, we did not find that the solutions
obtained using only an unpolarized GC model were stable or improved the image
quality.  The lack of polarized point sources as calibrators limits our
interpretive power for addressing some aspects of polarization leakage, which we
discuss in Section~\ref{sec:results}.

\begin{figure}
\centering
\includegraphics[scale=0.5]{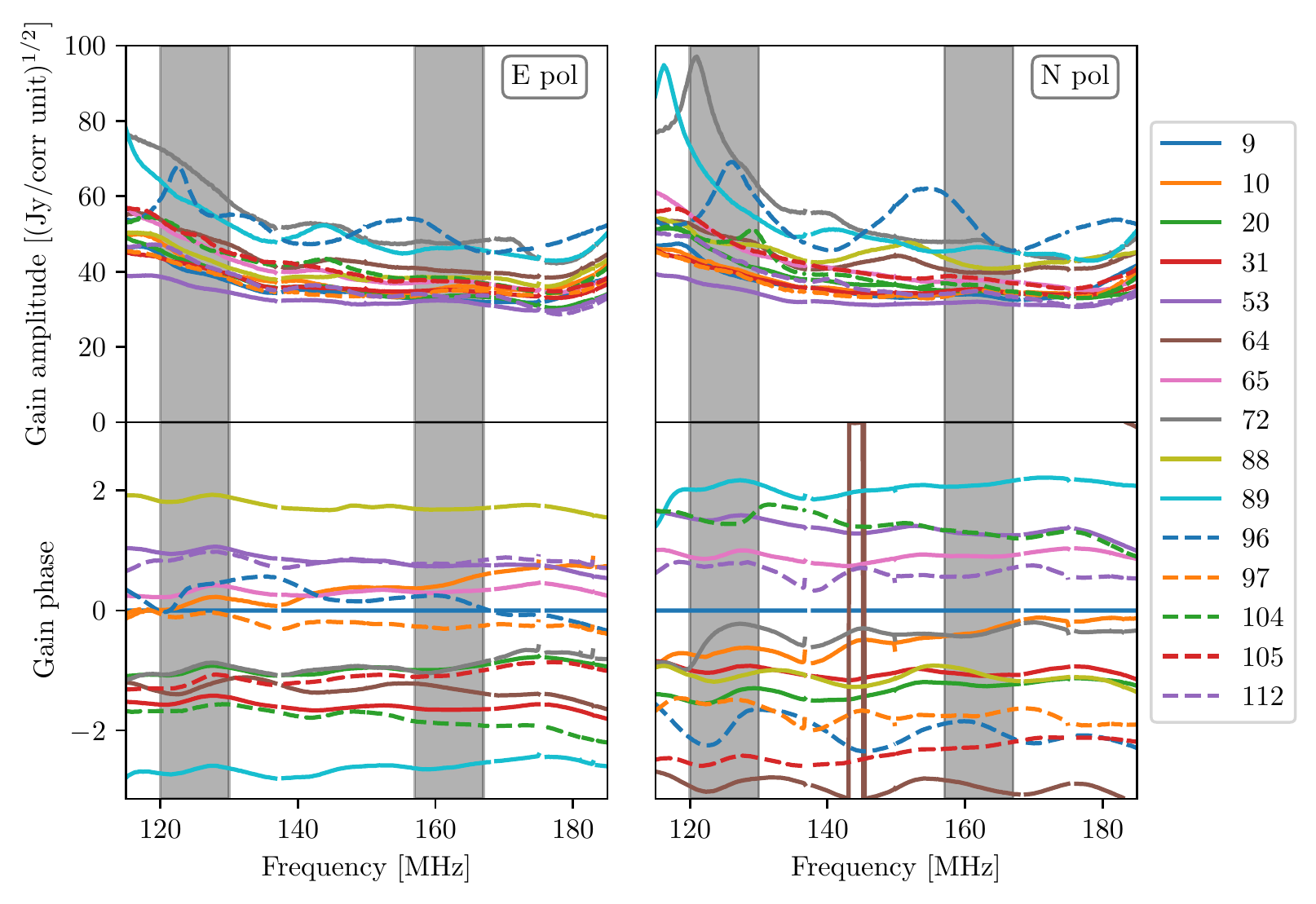}
\caption{Bandpass solutions obtained for both dipole orientations for all
  functioning antennas in the array on JD 2457548, and subsequently applied to
  all data.  Each antenna is marked by a different line color and style. Shaded
  regions indicate the effective sub-bands (the 10 MHz at the center of the 20
  MHz Blackman-Harris window) used for power spectrum analysis.  The phase is
  shown after the removal of the delay term.}
\label{fig:bandpass} 
\end{figure}

The calibration we have obtained serves to correct an initial large cable delay
per antenna, which aligns all of the power spectra at zero delay, and sets the
overall flux scale.  The resulting complex antenna-based gains are shown in
Figure~\ref{fig:bandpass}.  The gain amplitudes are clearly very similar in
shape, with one outlier, and they cluster with 25\% of each other.  After
removing the phase due to the delay term, the resulting phases show only small
variations around their mean.  The derived bandpasses are clearly spectrally
smooth, and thus, even if there are errors, we expect that they will not add
additional spectral structure to the power spectrum (see Section
\ref{subsec:pspec}).  These gains were applied to all 8 nights of
observations. It was found that this produced smaller day-to-day calibration
variability than calibrating each day separately to the GC.  An estimate of the
remaining variation is discussed in Section \ref{sec:Noise}.

\begin{figure}
\centering
\includegraphics[scale=0.38]{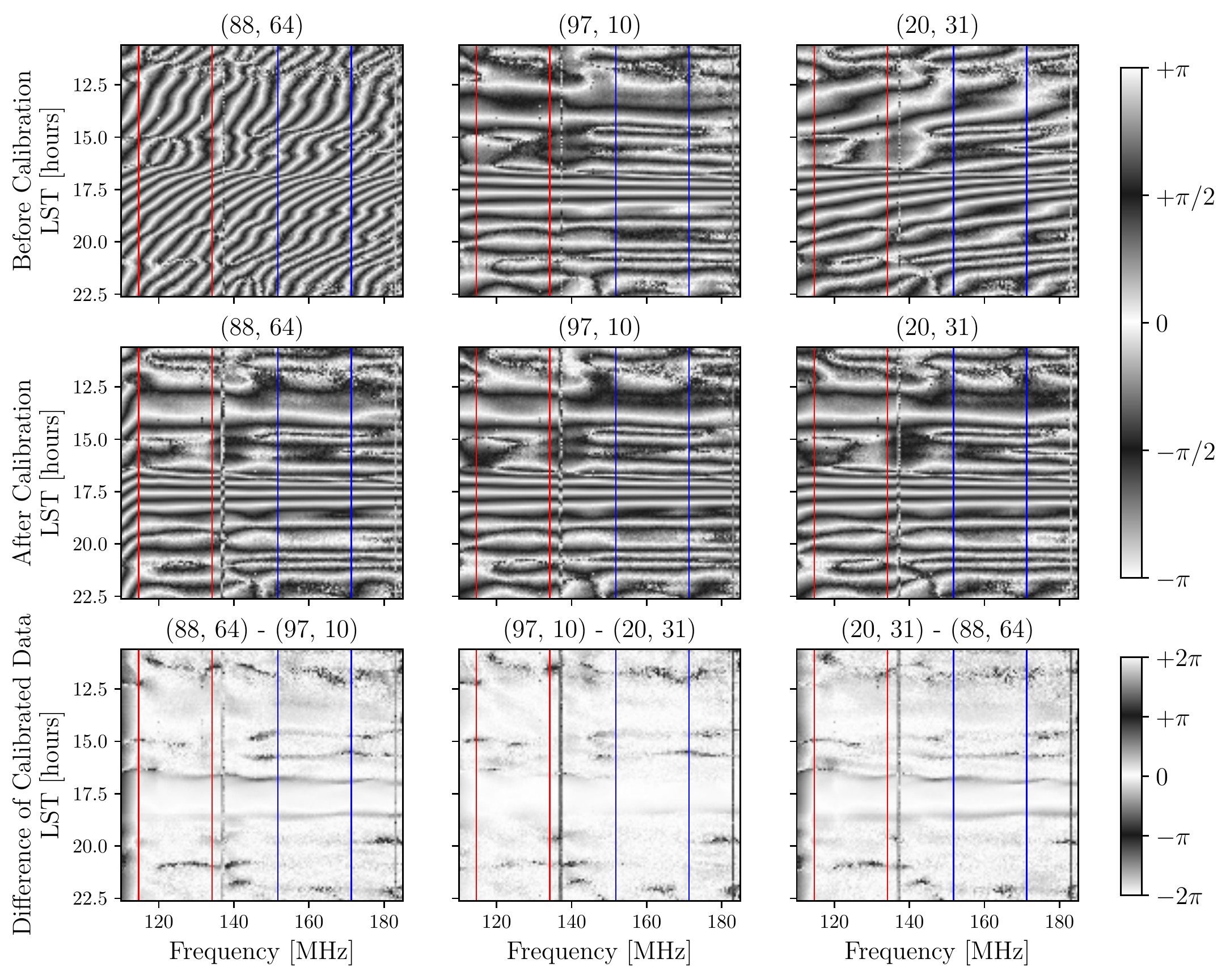}
\caption{ The effect of calibration on the phases of visibilities from three
  redundantly-spaced 14.7\,m baselines; \textit{nn} polarization.  The antenna
  numbers refer to those given in Figure \ref{fig:antpos}.  The color scale is
  cyclic; black is $\pm\pi/2$ and white is 0 and $\pm\pi$. The extend of the low
  band is indicated with red lines and the high band with blue.  \textit{Top
    row}: before calibration; \textit{Middle row}: after calibration;
  \textit{Bottom row}: the three pairs of differences of the calibrated phases.
  Note that the agreement between baselines is excellent near the Galactic
  Center but shows significant differences at some other times.  }
\label{fig:phasecal}
\end{figure}

Figure~\ref{fig:phasecal} shows the effect of calibration on the visibilities of
three nominally redundantly-spaced baselines.  Shown in that figure are the
phases of three $V^{nn}$ visibilities from 14.7\,m baselines before and after
calibration. There were no shared antennas between the visibilities shown. The
qualitative agreement is obvious, providing a consistency check on the
solutions, and showing our sky-based model achieves redundancy without assuming
it.  However, small-scale variations between baselines are still seen.  This not
unexpected; the antennas are likely to be non-identical, and we have evidence
based on closure phase (which is insensitive to calibration errors) that
redundant baselines do not see the sky identically \citep{carilli18}.

\begin{figure*}
\centering
\includegraphics[width=0.7\textwidth]{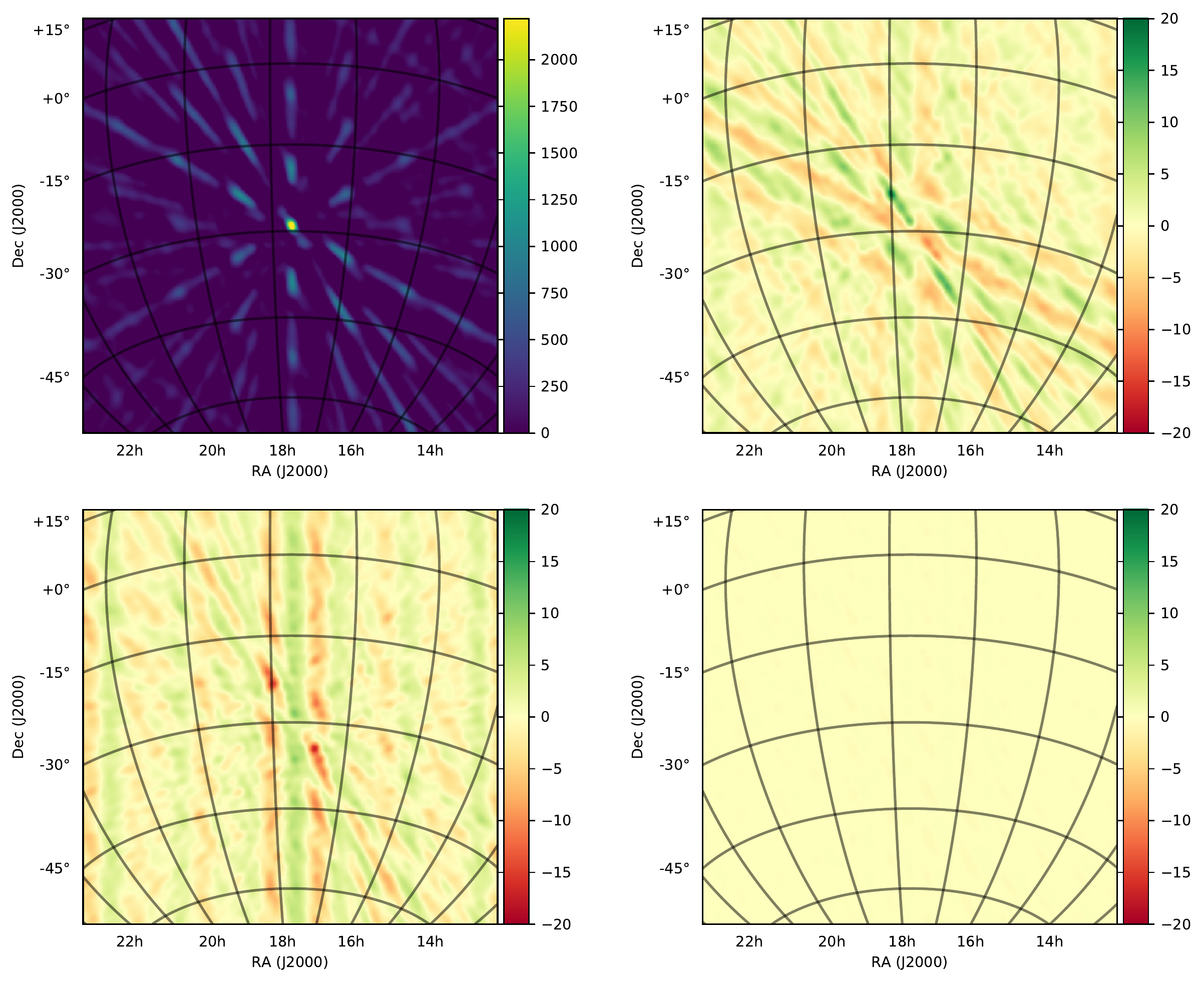}
\vspace{-0.1in}
\par\noindent\rule{0.8\textwidth}{0.4pt}
\includegraphics[width=0.7\textwidth]{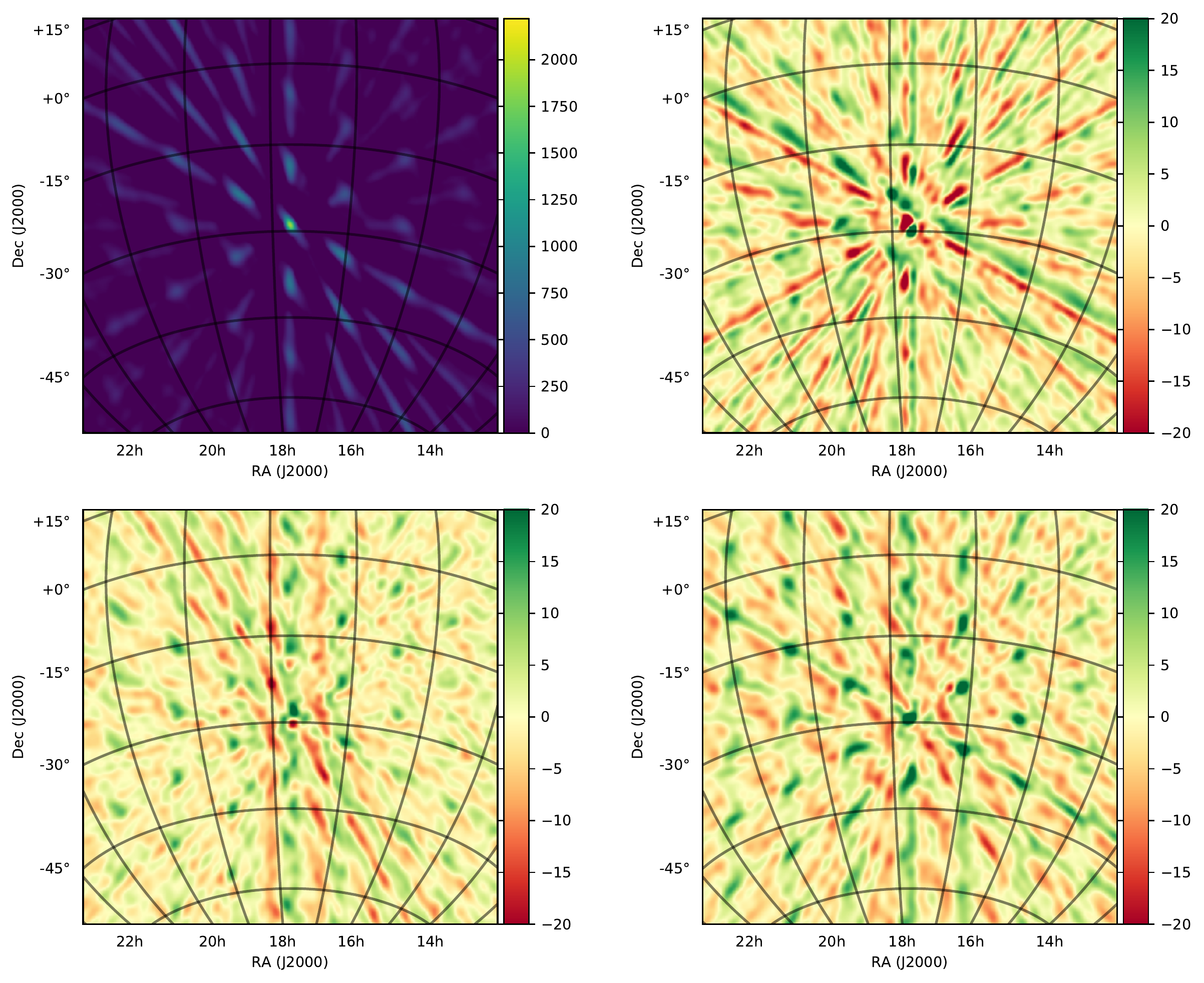}
\caption{Both sets of panels show multi-frequency synthesis pseudo-Stokes images
  of I, Q, U and V visibilities (\textit{top left, top right, lower left, lower
    right}) for the Galactic Center (our calibration source) at transit.  Note
  the field-of-view show is about 60\arcdeg across.  A Briggs-weighting with
  robustness 0 was used when gridding into the image plane.  No deconvolution
  was performed.  The colorbar is in units of Jy/Beam.  A separate color scale
  is used for Stokes I for suitable dynamic range in the polarized fluxes; note
  that the color scales differ by a factor of 100.  \textit{Above}: Simulation,
  where only a Stokes I sky was used; any polarized power is due to
  direction-dependent polarization leakage (see Section~\ref{subsec:DD-Leak}).
  \textit{Below}: Multi-frequency synthesis pseudo-Stokes images formed from
  observed visibilities on JD 2457548.}
\label{fig:GCimage}
\end{figure*}

In Figure~\ref{fig:GCimage}, we show images formed from the simulated
pseudo-Stokes visibilities (top panels) and our observations (bottom panels).
These are multi-frequency synthesis images, where we used all unflagged
frequencies on either side of the band edges, from 115 MHz to 188 MHz.  The
primary beam has not been deconvolved.  All images shown were produced using the
same four minute interval used for calibration.  Note that at HERA's latitude
the Galactic Center transits 2\arcdeg\ north of zenith, while the HERA primary
beam has a FWHM of $\sim10\arcdeg$ at 150\,MHz \citep{Neben.16}. For the
simulated visibilities, we flagged the same antennas and channels as in the
data. As expected for a compact array, the Stokes I images capture only a
low-resolution view of the Galactic Center. The simulated and observed
visibilities form remarkably similar images in Stokes I, and Q and U clearly
share features in common, due to leakage from I to Q and U through the primary
beam (recall that the simulations are unpolarized).  In Stokes V, the simulated
map has a significantly smaller amplitude of features compared to the actual
image generated from data.  The presence of emission at the location of the
Galactic Center not due to primary beam leakage is consistent with
direction-independent gain errors at the few percent level in amplitude (for
Stokes Q) and D-terms at $\sim 1\%$ relative to the diagonal gains (for Stokes
V) \citep{TMS}.  Note that the Stokes U image is broadly consistent with a large
fraction of power coming from I leakage through the primary beam, though there
is some additional power as well.  We consider the implications for the power
spectrum in Section~\ref{sec:results}.

\subsection{Forming power spectra}
\label{subsec:pspec}

Power spectra were formed in a fashion similar to the method used in
\cite{Pober13} and \cite{Kohn16}.  The actual implementation of the code is
available as part the {\tt GitHub} HERA-Team repository
\href{https://github.com/HERA-Team/hera_pspec}{\underline{{\tt hera\_pspec}}}.
We briefly review the method here.

\cite{Parsons.12a} define the \textit{delay transform} as the Fourier transform
of a visibility for baseline $ij$ and pseudo-Stokes parameter $P$ along the
frequency axis
\begin{equation}
\label{eq:DelaySpectrum}
\tilde{V}_{ij}^{P}(\tau, t) = \int {\rm d}\nu \tilde{V}_{ij}^{P}(\nu, t)e^{2\pi i \nu \tau}.
\end{equation}

We selected two relatively RFI-free 20 MHz sub-bands (Figure~\ref{fig:rfi}); 115
to 135 MHz and 152 to 172 MHz, henceforth referred to the ``low band'' and the
``high band'' in which to compute the power spectrum.  An extremely conservative
cut on RFI was used such that any integration which had RFI flagged in the 20
MHz sub-band was excluded from the analysis.  This cut was performed separately
for the two bands.  In each case, approximately 35\% of the data was retained
after this cut.
The bands were then multiplied by a Blackman-Harris window, centered on their
central frequencies, before Fourier transforming, in order to minimize
side-lobes. This windowing led to an noise-effective bandwidth of 10 MHz.  We
note that this bandwidth is appropriate for EoR analyses since the {\sc Hi}
signal is, to a reasonable approximation, coeval over the corresponding redshift
range \citep{Furlanetto06}.  However, this resolution (approximately 100 ns in
delay as compared to 194 ns for the longest baseline in this study) does limit
our ability to resolve certain features in the power spectrum.  We also note
that using a Blackman-Harris window will induce a correlation between adjacent
$\tau$ modes; this should be borne in mind when interpreting plots, as all delay
bins are plotted.

The power at each delay-mode and baseline can be represented in terms of their
respective Fourier components $k_{\parallel}$ and $k_{\perp}$
\citep{Parsons.12a, Nithya.15b}:
\begin{multline}
P(k_{\parallel},k_{\perp}) \approx | \tilde{V}_{ij}^{P}(\tau) |^2 \frac{X^2 Y}{\Omega_{pp} B_{pp}} \left(\frac{c^2}{2k_B\nu^2}\right)^2 , \\\\
k_{\parallel} = \frac{2\pi \nu_{\rm 21cm} H(z) 
}{c (1+z)^2}\tau, \\\\
k_{\perp} = \frac{2\pi}{D(z) \lambda} b\\
\end{multline}
for: cosomological bandwidth $B_{pp}$ and cosmological angular area of the beam
$\Omega_{pp}$, $\nu_{\rm 21cm}\approx$1420 MHz, baseline length $b$, wavelength
of observation $\lambda$, Hubble parameter $H(z)$, transverse comoving distance
$D(z)$ and redshift-dependent scalars X and Y \citep{Parsons.12b}. Note that the
angular area of the beam refers to the diagonal components of the Mueller
matrices shown in Figure~\ref{fig:mueller}. For further discussion of forming
polarized power spectra in $k$-space, refer to \cite{Nunhokee.17}.

To avoid a noise bias when forming the power spectrum, we cross-multiplied
consecutive integrations (each having independent noise), rephasing the zenith
angle of the latter to the former:
\begin{equation}
\label{eq:adj_time_ps}
 | \tilde{V}_{ij}^{P}(\tau, t) |^2 \approx | \tilde{V}_{ij}^{P}(\tau, t) \times \tilde{V}_{ij}^{P}(\tau, t+\Delta t)e^{i\theta_{ij,\rm zen}(\Delta t)}|
\end{equation}
where $\theta_{ij,\rm zen}(\Delta t)$ was the appropriate phasing for baseline
$ij$ and $\Delta t = 10.7$ seconds.

Pseudo-stokes power spectra were formed for each pair of integrations, for every
baseline, according to Equation \ref{eq:adj_time_ps}.  Power spectra from
baselines of identical lengths were then averaged together for all observation
times over 8 days.
The resulting ``1D'' power spectra for each baseline length were then
concatenated to form a two-dimensional power spectrum (that is, arranged into
the ($k_{\perp}$, $k_{\parallel}$) plane).  Note that all averaging in this
study was performed {\it after} forming power spectra, {\it not} by averaging
visibilities; this incoherent averaging is non-optimal for achieving the
greatest sensitivity.  Future work will be able to test the features of the
polarized beam and foregrounds too much greater depth.

\section{Results}
\label{sec:results}

\begin{figure*}[h]
\centering
\includegraphics[scale=0.45]{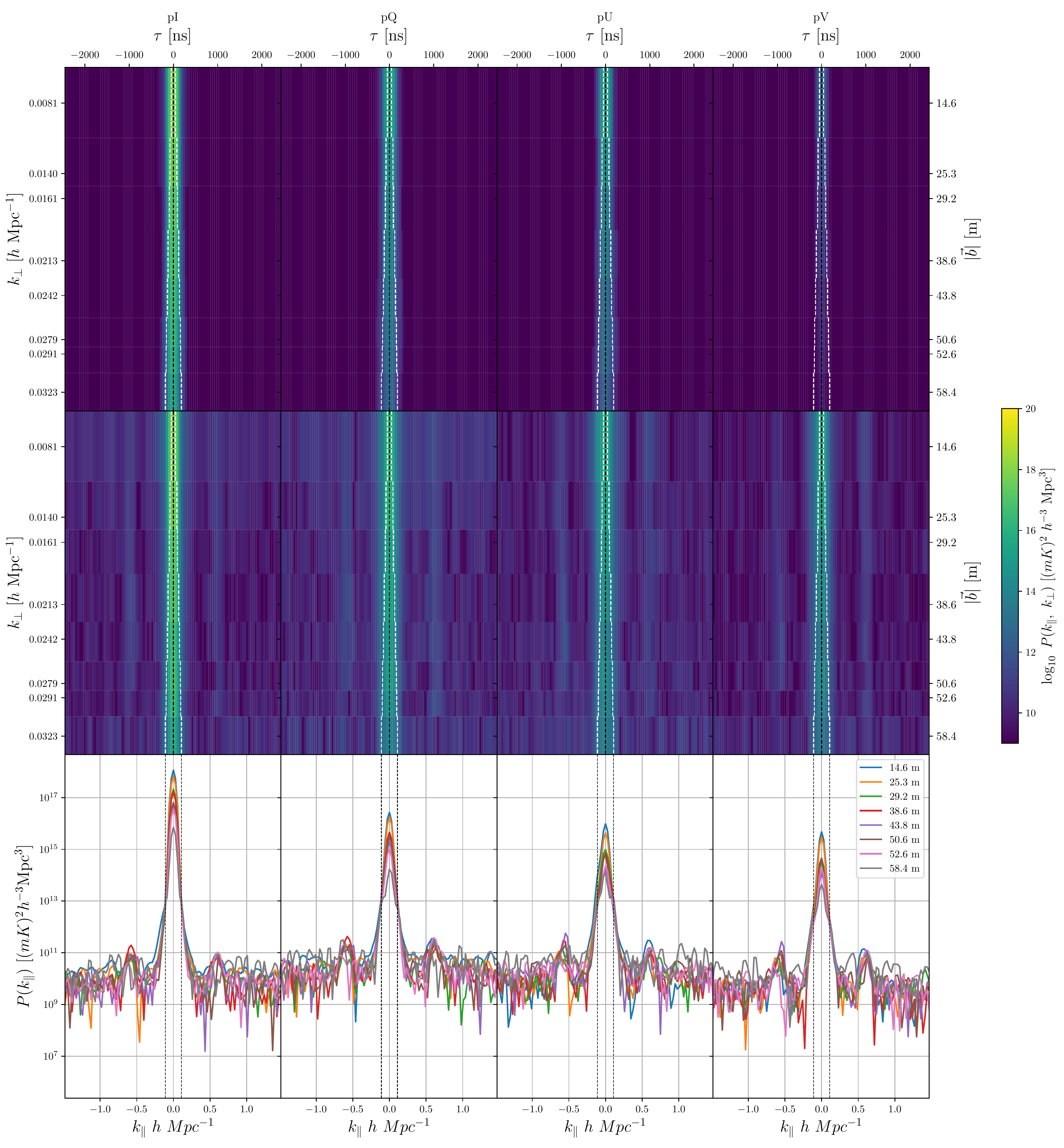}
\caption{Results from the high-band (157--167 MHz).  \textit{Top}: Simulated
  power spectra in Stokes I, Q, U and V, following the formalism in
  Section~\ref{sec:leak}.  No polarized sky model was used, so power in Stokes
  Q, U and V is only due to the direction-dependent beam leakage from Stokes I.
  No instrumental noise was included in the simulation. \textit{Middle}:
  Eight-day average power spectra from data. \textit{Bottom}: The same data as
  shown in the middle panel, but with each baseline length overlaid on one
  another to allow shared features to be more easily identified.  For the top
  and middle plots, white dotted lines indicate the boundary of the pitchfork
  and the EoR window for that baseline length. A black dotted line indicates the
  $k_{\parallel}=0$\,h/Mpc line.  In the bottom panel, dotted lines indicate the
  boundary of the wedge for the longest baseline only.  Delays in nanoseconds
  are indicated along the top, and the corresponding cosmological $k$ at the
  mean redshift along the bottom axis.}
\label{fig:pitchforks_highband}
\end{figure*}

\begin{figure*}[h]
\centering
\includegraphics[scale=0.45]{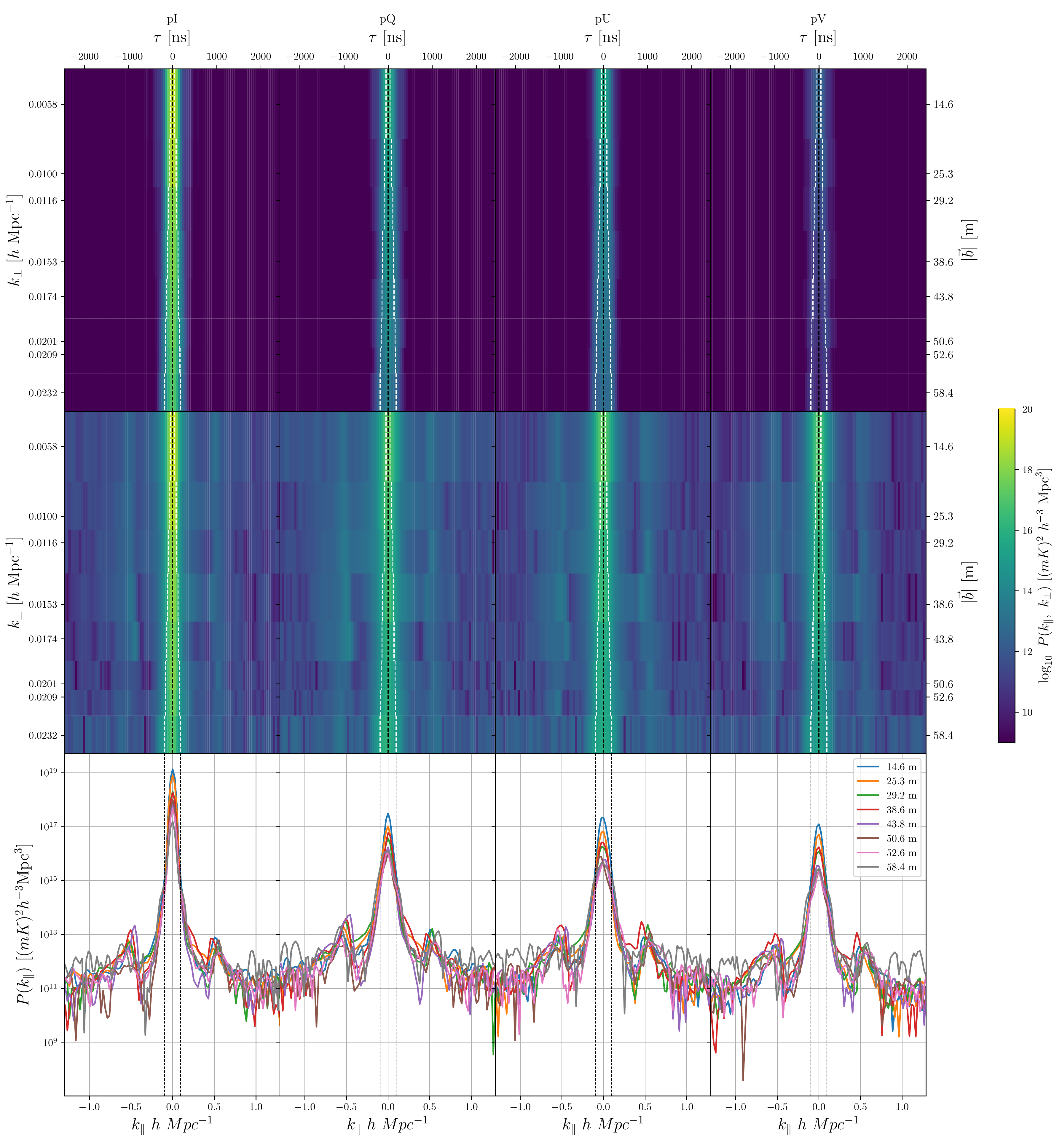}
\caption{Results from the low-band (120--130 MHz), arranged in the same format
  as Figure~\ref{fig:pitchforks_highband}.}
\label{fig:pitchforks_lowband}
\end{figure*}

The power spectra formed from the above procedure are shown for all
pseudo-Stokes parameters for the high and low bands in
Figures~\ref{fig:pitchforks_highband} and \ref{fig:pitchforks_lowband},
respectively.

\subsection{General features of the power spectra}
\label{subsec:general_features}

Several features of the power spectra are readily apparent.  The first is that
foreground emission appears in a relatively narrow band near $k_\parallel = 0$.
Another is that the the shape of the power spectrum as a function of
$k_\parallel$ is both sharply peaked and relatively featureless.  We note that a
similar study of 2D polarized power spectra in \citet{Kohn16}, PAPER
measurements showed a comparably ``filled'' region of Fourier space out to the
horizon delay (i.e., to directions corresponding to zenith angle
$\pm$90$^{\circ}$), with some supra-horizon leakage \citep[e.g.,][]{Pober13}
into the EoR window itself. The power spectra in
Figures~\ref{fig:pitchforks_highband} and \ref{fig:pitchforks_lowband} show
similar behavior, though in this case part of the reason is the low resolution
($\sim100$ ns; see Section \ref{subsec:pspec}) of the delay transform (Equation
\ref{eq:DelaySpectrum}) due to the small spectral bandwidth, and the small
horizon delay associated with the short baselines of the array.  Thus we are not
able to verify the prediction of \citet{Nithya.15b} and \citet{Neben.16} that
for dishes such as HERA, the region between a delay of about 50 ns (set by the
width of the antenna primary beam) and the horizon should be free of foreground
emission.  Similarly, an ``excess'' of power near the horizon delay, as
predicted by the same authors, is not observable, again due to the blurring of
features in $k_\parallel$ by the resolution.  Along the $k_\perp$ direction in
Stokes I, the amplitude declines as a function of $k_\perp$, as expected for
diffuse Galactic emission with a power law angular power spectrum (larger
fluctuation power on large scales).  This trend is also observed for the other
Stokes parameters as well.

A notable feature in the observational data is a peak in power above the noise
level at a delay of $\sim1000 \, \mathrm{ns}$, independent of the baseline
length or the frequency band.  There are $\sim 150 \, \mathrm{m}$ coaxial cables
connecting the HERA dishes to the correlator\footnote{This stage of the signal
  chain is only present in the commissioning array. Future HERA build-outs will
  transition to a different architecture using RF over fiber with long fiber
  runs to move this signal to even longer delays \citep{deBoer17}.} and we have
evidence that some of this power is due to a cable reflection at this stage of
the signal chain producing an alias of the foreground signal; see
\href{http://reionization.org/wp-content/uploads/2013/03/HERA39_H1C_cable_reflections_ewall-wice.pdf}{\underline{HERA
    Memo \#39}} \citep{hera_memo39}.  However, this signal should appear at a
delay corresponding to 1300 ns, or twice the propagation time in the coaxial
cable.  It appears that most of the signal present here is present at a smaller
delay, and its origin is not understood.

\subsection{Comparison to Simulations}

\begin{figure*}[h]
\centering
\includegraphics[width=0.9\textwidth]{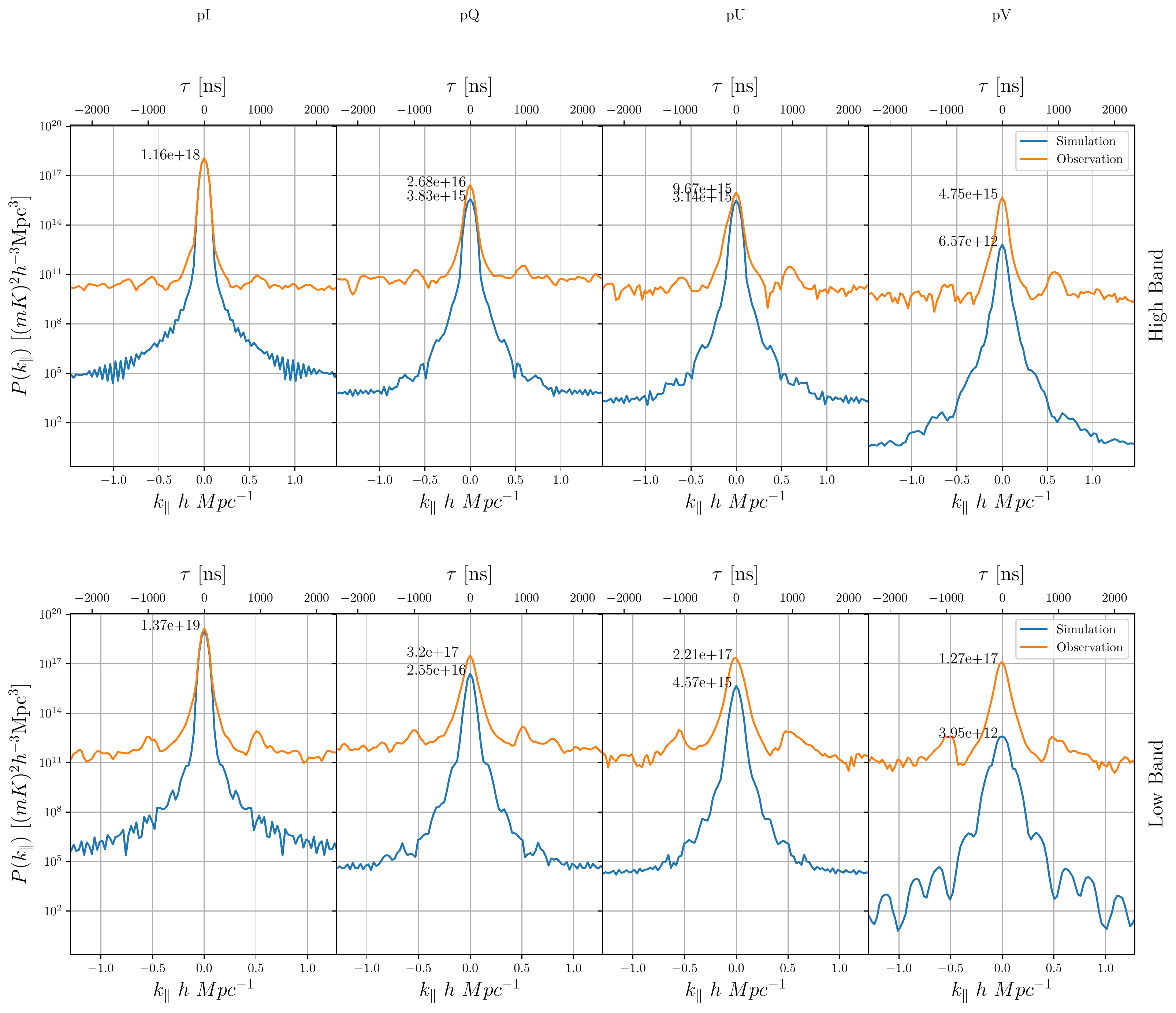}
\caption{Simulated and observed power as a function of $k_{\parallel}$ for the
  shortest baseline (14.6\,m). \textit{Left to right}: pseudo-Stokes I, Q, U and
  V; \textit{above}: the high band; \textit{below}: the low band. The
  simulations were noiseless and used an unpolarized sky model. The agreement
  with Stokes I is excellent in the high band, and consistent with an absolute
  calibration accuracy of $\sim$20\% for the low band.  The agreement between
  the simulations and the data for the other pseudo-Stokes parameters is poor,
  as discussed in the text, likely due to a combination of calibration errors,
  particularly for Stokes V, and actual polarized emission.}
\label{fig:bl0_cuts_vs_sim}
\end{figure*}

\begin{figure*}[h]
\centering
\includegraphics[width=0.9\textwidth]{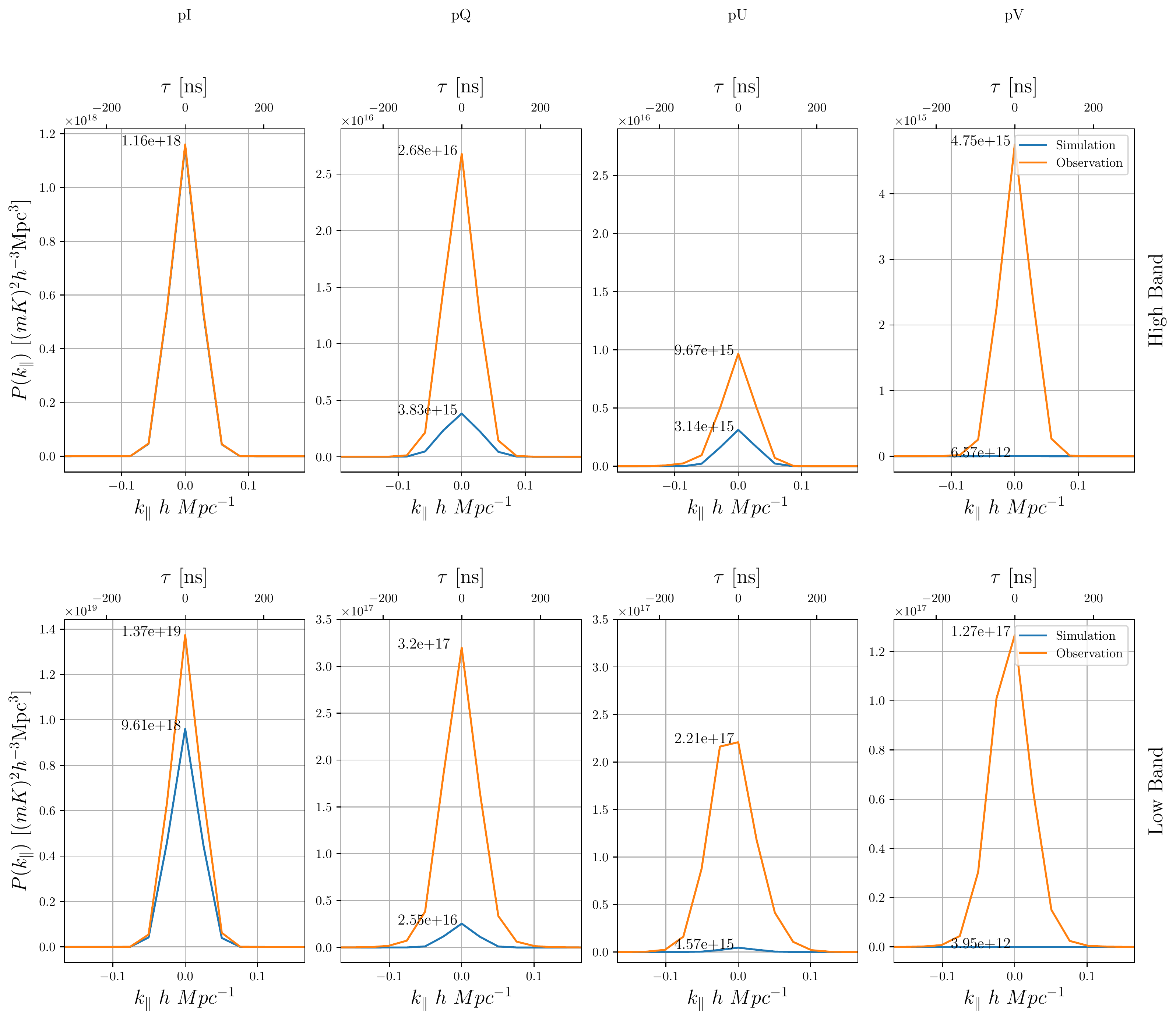}
\caption{A zoom in on Figure \ref{fig:bl0_cuts_vs_sim} showing just the
  $k_\parallel$ values near the wedge, and with a linear scale in
  $P(k_\parallel)$.  Note that the scale changes for each Stokes for each
  parameter, except that Q and U are set to have the same scale in each row.}
\label{fig:bl0_cuts_vs_sim_zoom}
\end{figure*}

In Figure~\ref{fig:bl0_cuts_vs_sim}, we show a direct comparison between the
power spectra of the data and the simulations for the shortest baseline length
for all Stokes parameters for both bands. Figure~\ref{fig:bl0_cuts_vs_sim_zoom}
shows a zoom-in of the same data close to the wedge. Recall that the simulations
include only a Stokes I sky component and no simulated calibration errors, so
the signal in the simulated pseudo-Stokes Q, U, and V power spectra is due
solely to wide-field beam leakage (Figure~\ref{fig:mueller}). The Stokes I model
is the diffuse emission of the GSM, which should be accurate at the scales
probed by a 14.7\,m baseline, as the model includes scales down to 1\arcdeg.
The power spectra of the simulated data have been computed in the same way as
the data.

In comparing to the foreground simulations, there are two things to notice.  The
first is the {\it isolation}, or the degree to which the foregrounds remain
within the range of $k_\parallel$ defined by their intrinsic smoothness and by
the mode-mixing of the interferometer, and which can be characterized by the
width of the foregrounds in $k_\parallel$ (or $\tau$) space.  The second is the
{\it dynamic range}, defined here as the ratio between the $k_\parallel=0$ peak
and the smallest value of the power spectrum.  This smallest value may be
limited either by noise (in the case of the data) or by the Blackman-Harris
window (for the noise-free simulations).  We note that the simulations provide a
reasonable standard by which to judge both isolation and dynamic range: in the
absence of any systematic effects, the width of the foregrounds in delay space
cannot be narrower than that captured by the simulated foreground spectral
structure and instrument mode-mixing, and the dynamic range is as expected from
the window function.  Thus over a range of about 8 orders of magnitude in the
Stokes I power spectrum, limited by the noise in the data, the isolation in the
data agrees well with the simulation, arguing that the calibrated data (at the
current noise level) do not have significant spectral structure beyond that
intrinsically present.  It is worth pointing out there is some evidence for
broadening of the range in $k_\parallel$ near the noise floor.  For the the
other pseudo-Stokes spectra, the isolation is worse: the power spectrum of the
data is noticeably wider than the simulation.

Referring to Figure \ref{fig:bl0_cuts_vs_sim_zoom}, calibrating the raw data to
the simulation as described in Section \ref{subsec:cal} reproduces the total
power in I very well in the high band, and leads to a total power $\sim$40\%
higher than the simulation in the low band. This discrepancy is consistent with
the overall amplitude of the calibration derived from the GC disagreeing by
$\sim$20\% in the low band. It is not clear whether this is due to errors in the
GSM or the primary beam model at low frequencies.  In the analysis that follows,
we have increased the amplitude of the simulated power spectrum in the low band
to agree with the real data because we are primarily concerned with the relative
power between different pseudo-Stokes parameters; this rescaling allows for a
more even treatment of the two bands.

In the high band analysis, the beam leakage modeled in the simulation is of
roughly equal magnitude for both Q and U; however, the data show a markedly
stronger pseudo-Q than U. Because there is no strong reason to believe the sky
is highly anisotropic between Q and U over the RA included here, this difference
argues for another interpretation. The amplitude difference between pseudo-Q and
U is $\sim$10\% in the visibilities (rather than the power spectra), which can
be accounted for by a $\sim$5\% relative miscalibration in the antenna-based
gain amplitudes between the E and N polarized feeds of the antenna. The pseudo-Q
visibilities are constructed by differencing the calibrated \textit{nn} and
\textit{ee} visibilities, which can make relative amplitude differences between
the two polarizations more pronounced. Thus, assuming the excess power in
pseudo-Q can be attributed to this difference in gain amplitude between the two
polarized feeds, we can use pseudo-U as a measurement of the excess polarized
power on the sky not accounted for in the beam leakage from the
simulations. This interpretation leads to a combined fractional polarization of
$\sim$10\%. This result is on the high end of the range of measurements of
$\sim 1.6 - 4.5$\% fractional polarization at 150\,MHz on large-scales from
\cite{Jelic.15} and \cite{Lenc.16}.  We note that whatever the interpretation of
the linear pseudo-Stokes spectra, they do not provide strong evidence for high
rotation measure emission, which would be present at $k_\parallel > 0.1$ for
$RM > 10~\text{rad m}^{-2}$ in the high band \citep{Moore17}, although noise
prevents probing levels as deep as those in \cite{Asad18}.

As with pseudo-Q and U, pseudo-V has excess power in the measured power spectra
compared to the reference simulations. Measurements in the literature do not
suggest a significant amount of large-scale circularly polarized emission at the
frequencies measured, and so this excess power is most likely due to
miscalibration of the instrument. Relative phase errors in the gain solutions of
the cross-polarized instrumental visibilities can lead pseudo-U to leak into
pseudo-V. However, the measured amplitudes of the two power spectra would imply
that this systematic error in phase angle must be order $\pi/2$, or that these
phase errors are nearly maximally rotated with respect to the correct value. A
more plausible explanation for this excess power in pseudo-V is direction
independent leakage through the $D$-terms of the Jones matrix, which leak power
from pseudo-I \citep{TMS}. As discussed in Sec.~\ref{subsec:cal}, $D$-terms were
entirely neglected for the calibration performed in this analysis. Assuming that
these $D$-terms are the source of the leaked power from pseudo-I to pseudo-V,
their amplitude relative to the diagonal elements of the Jones matrix would be
$\sim$3\%. This is similar to $D$-term levels from other low frequency
instruments such as MWA-32, which was found to have $\sim$2\% $D$-terms
(G. Bernardi, private communication).

The low band measurements tell a similar story, with the notable difference that
the simulated pseudo-Q and U power spectra exhibit a greater discrepancy than
the high band measurements.  This difference may be attributable to the general
difficulty in accurately calibrating the low band due to model and beam
uncertainties.  Performing the same analysis as for the high band, we again find
that the inferred polarization fraction from the pseudo-U visibilities is
$\sim$10\%, and the excess power in pseudo-Q can be accounted for by gain
amplitude errors of $\sim$5\%. The relative amplitude of the $D$-terms to
account for the excess power in pseudo-V are closer to 5\%, which is slightly
higher than the high band but still plausible given the behavior of similar
instruments. These results suggest that future precision calibration efforts
should include analysis of $D$-terms in order to accurately model the
instrumental effects on the measured visibilities.

\subsection{Noise Levels}
\label{sec:Noise}

One estimate of the system temperature of the observations was formed from the
calibrated values of the auto-correlations.  These were compared against the
values obtained from the simulation.  Over the RA range observed, which was
heavily weighted towards the Galactic Center and much of the Galactic Plane, the
system temperature estimated in this way was 1230 K for the high band and 4000 K
for the low band, which was consistent with the simulated autocorrelations
\citep[][also see the public
\href{http://reionization.org/wp-content/uploads/2017/04/HERA19_Tsys_3April2017.pdf}{\underline{HERA
    Memo \#16}}]{deBoer17}.

\begin{figure}
\centering
\includegraphics[width=0.9\columnwidth]{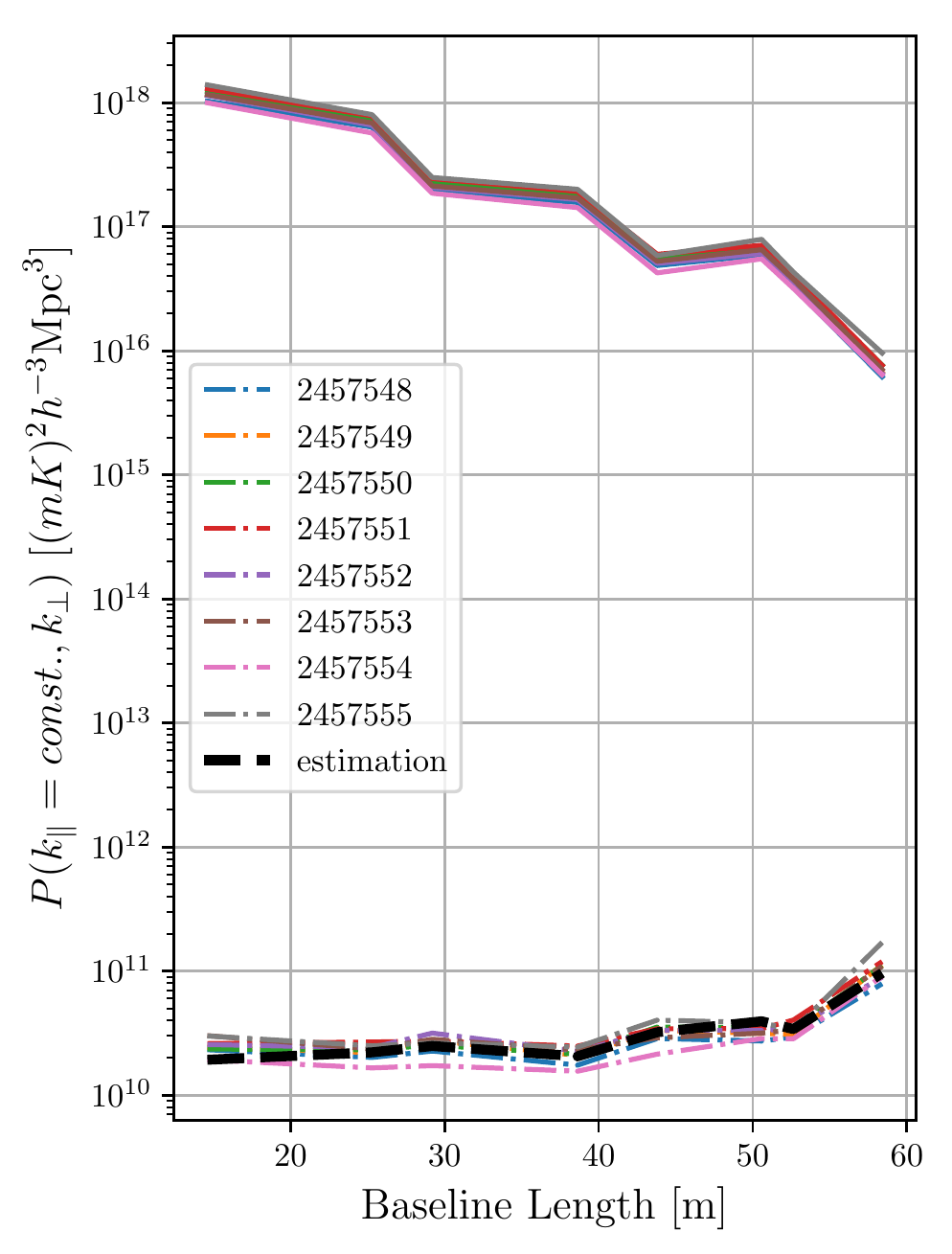}
\caption{High band power as a function of baseline length for $k_{\parallel}=0$
  h/Mpc (solid lines) and an average value over the white noise
  ($|k_{\parallel}|>1$ h/Mpc; dot-dashed lines) for pseudo-Stokes I on each JD
  of observation. The black dashed line represents estimated power spectrum
  noise given a system temperature of 2400\,K.  A very similar relationship
  obtains for the low band (not shown), but with a higher system temperature of
  7000 K.  Note that the increase of the noise level with baseline length is
  correctly modeled, as the compact array has has a decreasing number of
  baselines which are averaged together in a given $(k_{\parallel},k_{\perp})$
  bin.}
\label{fig:highband_cuts_per_day}
\end{figure}

The system temperature was converted to a noise level in the power spectrum
according to the formalism in \cite{Parsons.12a}, with the inclusion of a
baseline-number dependence:
\begin{equation}
P_{\rm Noise}(k) \approx \frac{1}{2\Delta t \sqrt{N_{\rm LST}} N_{\rm days} N_{\rm bl}} X^2 Y B_{\rm NE} \Omega_{\rm eff} T_{\rm sys}^2.
\end{equation}
In the above equation, $\Delta t$ is integration time, $N_{\rm LST}$ is the
number of LST hours used per day (12 hours), $N_{\rm bl}$ is the number of
baselines (which differed per $k_{\perp}$ bin), $X$ and $Y$ are cosmological
scalars defined in \cite{Parsons.12a}, $B_{\rm NE}$ is the noise equivalent
bandwidth and $\Omega_{\rm eff}$ is the effective beam area, as defined in
\cite{Parsons14}.  Figure~\ref{fig:highband_cuts_per_day} shows power as a
function of baseline length for $k_{\parallel}=0$ h/Mpc (solid lines) and the
average over $|k_{\parallel}|>1$ h/Mpc (dot-dashed lines).  Though not shown in
Figures \ref{fig:pitchforks_highband} and \ref{fig:pitchforks_lowband}, the
frequency sampling of the instrument is sensitive to delays up to 5000 ns, and
this region defines our white noise level.  To match the observed noise level of
the power spectrum in Stokes I, high band, required an assumed system
temperature of 2400 K, higher than that expected based on the sky and beam model
from the simulations, or from the measured calibrated autocorrelations.  The
cause of this excess is not understood.  Figure ~\ref{fig:highband_cuts_per_day}
also shows that both the noise and calibration of the instrument were stable at
the 20\% level in the visibilities (50\% in the power spectrum) over the 8 days
presented here.

\section{Conclusions}
\label{conc}

In this work we have investigated the polarization response of the HERA-19
commissioning array, both in imaging and particularly in the power spectrum
domain.  We find that a simple image-based calibration based on the unpolarized
diffuse emission of the Global Sky Model (GSM) has a spectrally smooth structure
and achieves qualitative redundancy between the nominally-redundant baselines of
the array.  We are able to calibrate the data based on the Galactic Center
observations to the GSM with an accuracy of about 10\%, and about 20\% variation
from day-to-day.

Forming power spectra in all pseudo-Stokes parameters, we show that we achieve
isolation of the foregrounds in Stokes I as expected due to their intrinsic
spectral smoothness, the modeled instrument chromaticity, and the calibration,
limited in dynamic range by the noise.  Excess power at a delay of $\sim1000$ ns
is seen in all polarizations which may in part be due to cable reflections, but
is not fully explained.  Excess power is also seen in the power spectra of the
linear polarization Stokes parameters which is not easily attributable to
leakage via the primary beam, and results from some combination of residual
calibration errors and actual polarized emission.  Finally, Stokes V is found to
be highly discrepant from the expectation of zero power, likely due to the lack
of calibration of off-diagonal Jones matrix (``$D$'') terms.

The results presented here are necessarily preliminary, and point in obvious
directions for improvements in the quality of calibration, particularly the
polarized calibration, which are currently being prepared by the HERA
collaboration.  Deeper integrations in the power spectrum will probe the
structure of the foregrounds and instrument response to a higher dynamic range
and over a wider range in $k_{\perp}$-modes as more antennas are added, allowing
a more thorough characterization of the wedge shape.  A build-out of HERA to 350
antennas with a new broad-band feed and completely new electronics chain is now
underway \cite{deBoer17}, with strong quality-assurance efforts informed in part
by this analysis.

\acknowledgements

This material is based upon work supported by the National Science Foundation
under Grant Nos. 1440343 and 1636646, the Gordon and Betty Moore Foundation, and
institutional support from the HERA collaboration partners.

HERA is hosted by the South African Radio Astronomy Observatory, which is a
facility of the National Research Foundation, an agency of the Department of
Science and Technology.

Much of this study was undertaken during the inaugural CAMPARE-HERA Astronomy
Minority Partnership (CHAMP), funded under the NSF grants.

SAK is supported by a University of Pennsylvania SAS Dissertation Completion
Fellowship.
JEA acknowledges support from NSF CAREER award \# 1455151.
CDN is supported by the SKA SA scholarship program.
AL acknowledges support from a Natural Sciences and Engineering Research Council
of Canada (NSERC) Discovery Grant and a Discovery Launch Supplement, as well as
the Canadian Institute for Advanced Research (CIFAR) Azrieli Global Scholar,
Gravity and the Extreme Universe Program.
GB acknowledges support from the Royal Society and the Newton Fund under grant
NA150184. This work is based on the research supported in part by the National
Research Foundation of South Africa (grant No. 103424).
JSD acknowledges the support of the NSF AAPF award \# 1701536 and the Berkeley
Center for Cosmological Physics.
MJK is supported by the NSF under project number AST-1613973
Parts of this research were supported by the Australian Research
Council Centre of Excellence for All Sky Astrophysics in 3 Dimensions
(ASTRO 3D), through project number CE170100013.

We thank the anonymous referee for many helpful suggestions.

\software{This research made use of Astropy, a community-developed core Python
  package for Astronomy \citep{astropy13}; CASA \citep{casa}; pyuvdata
  \citep{pyuvdata}; pygsm \citep{pygsm} }

\clearpage

\bibliography{mybib}
\bibliographystyle{apj}

\end{document}